\documentclass[journal,12pt,onecolumn,draftclsnofoot,a4paper,]{IEEEtran}
\IEEEoverridecommandlockouts
\usepackage{amsmath,amssymb,amsfonts}
\usepackage{algorithmic}
\usepackage{graphicx}
\usepackage{textcomp}
\usepackage{xcolor}
\usepackage[textsize=tiny]{todonotes}

\usepackage{cite}
\usepackage{xcolor}
\usepackage{url}
\usepackage{hyperref}
\usepackage{comment}
\usepackage{theorem}
\usepackage{subcaption}
\usepackage{calligra}
\usepackage{array}

\newcommand{\6}{\mathbf }

\usepackage{tikz}
\usetikzlibrary{arrows,automata, positioning}
\usetikzlibrary{calc, chains,
                fit,
                positioning,
                shapes}
\usetikzlibrary{patterns}       

\usepackage{pgfplots}
\usepackage{dashrule}

\usetikzlibrary{matrix,fadings,decorations.pathmorphing,intersections}
\usepgfplotslibrary{fillbetween}
\pgfdeclarelayer{bg}
\pgfsetlayers{bg,main}
\makeatletter
\tikzset{use path/.code=\tikz@addmode{\pgfsyssoftpath@setcurrentpath#1}}
\makeatother

\usetikzlibrary{automata,arrows,positioning,calc}
\definecolor{myblue}{RGB}{0,0,205}
\definecolor{mygreen}{RGB}{80,160,80}
\definecolor{myred}{RGB}{178,34,34}
\definecolor{mygray}{RGB}{211,211,211}
\definecolor{myyellow}{rgb}{1.0, 0.75, 0.0}
\definecolor{mygreen2}{rgb}{0.4, 0.69, 0.2}
\definecolor{myred2}{rgb}{0.89, 0.15, 0.21}
\definecolor{myblue2}{rgb}{0.0, 0.47, 0.75}

\usepackage[boxed, linesnumbered, noend]{algorithm2e}

\SetCommentSty{mycommfont}

{\theorembodyfont{\rmfamily}
   \newtheorem{Proper}{{\textbf Property}}}
{\theorembodyfont{\rmfamily}
   \newtheorem{The}{{\textbf Theorem}}}
{\theorembodyfont{\rmfamily}
   }
{\theorembodyfont{\rmfamily}
   \newtheorem{Lem}{{\textbf Lemma}}}
{\theorembodyfont{\rmfamily}
   \newtheorem{Cor}{{\textbf Corollary}}}
{\theorembodyfont{\upshape}
   }
{\theorembodyfont{\upshape}
   \newtheorem{Exa}{{\textbf Example}}}
{\theorembodyfont{\upshape}
   }
{\theorembodyfont{\upshape}
   \newtheorem{Rem}{{\textbf Remark}}}

\usetikzlibrary{patterns}
\usepackage{scalerel}

\usepackage[acronym,shortcuts]{glossaries}
\pgfplotsset{compat=1.18}

\newacronym{AWGN}{AWGN}{Additive White Gaussian Noise}
\newacronym{BPSK}{BPSK}{Binary Phase Shift Keying}
\newacronym{CER}{CER}{Codeword Error Rate}
\newacronym{LDPC}{LDPC}{Low-Density Parity-Check}
\newacronym{RC-LDPC}{RC-LDPC}{Rate-Compatible \ac{LDPC}}
\newacronym{PRC-LDPC}{PRC-LDPC}{Primitive \ac{RC-LDPC}}
\newacronym{IoT}{IoT}{Internet of Things}
\newacronym{RCC}{RCC}{Row-Column Constraint}
\newacronym{RU}{RU}{Richardson-Urbanke}
\newacronym{LFSR}{LFSR}{Linear Feedback Shift Register}
\newacronym{SPA}{SPA}{Sum-Product Algorithm}
\newacronym{BP}{BP}{Belief Propagation}
\newacronym{MS}{MS}{Min-Sum}
\newacronym{NMS}{NMS}{Normalized Min-Sum}
\newacronym{QC-LDPC}{QC-LDPC}{Quasi-Cyclic \ac{LDPC}}
\newacronym{LLR-SPA}{LLR-SPA}{Log-Likelihood Ratio \ac{SPA}}

\pgfplotsset{compat=1.17}
\begin{document}

\title{Rate-compatible LDPC Codes based on Primitive Polynomials and Golomb Rulers \thanks{The material in this paper has been presented in part at the 2022 61st FITCE International Congress Future
Telecommunications: Infrastructure and Sustainability (FITCE), Rome (Italy) \cite{Battaglioni2022Fitce}, and at the 2023 IEEE International Conference on Communications (ICC), Rome (Italy) \cite{Battaglioni2023ICC}.}
}

\author{Massimo Battaglioni, Marco Baldi, Franco Chiaraluce and Giovanni Cancellieri\\

\IEEEauthorblockN{
\textit{Dipartimento di Ingegneria dell'Informazione}
\\{Università Politecnica delle Marche, Ancona (60131), Italy }
\\email:\{m.battaglioni, m.baldi, f.chiaraluce, g.cancellieri\}@univpm.it}}

\maketitle

\begin{abstract}
We introduce and study a family of rate-compatible Low-Density Parity-Check (LDPC) codes characterized by very simple encoders. The design of these codes starts from simplex codes, which are defined by parity-check matrices having a straightforward form stemming from the coefficients of a primitive polynomial. For this reason, we call the new codes Primitive Rate-Compatible LDPC (PRC-LDPC) codes. 
By applying puncturing to these codes, we obtain a bit-level granularity of their code rates.
We show that, in order to achieve good LDPC codes, the underlying polynomials, besides being primitive, must meet some more stringent conditions with respect to those of classical punctured simplex codes. We leverage non-modular Golomb rulers to take  the new requirements into account. We characterize the minimum distance properties of PRC-LDPC codes, and study and discuss their encoding and decoding complexity. Finally, we assess their error rate performance under iterative decoding. 
\end{abstract}


\section{Introduction}
\label{introduction}

\Ac{LDPC} codes are a family of error correcting codes widely employed in modern communication systems, due to their ability to provide excellent error rate performance with relatively low decoding complexity. Recently, \ac{RC-LDPC} codes, introduced in \cite{Ha2004}, have gained increasing attention, since they offer the flexibility required by many modern applications. In fact, these codes can support different code rates by using a single code design, as for the wider class of general rate compatible codes \cite{DAVIDA1972}. Such a flexibility can represent a significant advantage, for example, in modern radio and wireless communications like the fifth generation (5G) or the sixth generation (6G) of mobile communications, since the underlying transmission systems need to support a wide range of data rates and channel conditions to meet the requirements of the various application scenarios \cite{3GPP}. 
Another advantage of \ac{RC-LDPC}  codes is their ability to allow for low encoding and decoding complexity. In fact, due to the fact that all the obtainable codes can be constructed starting from a common structure (e.g., a base matrix), their encoding and decoding procedures can be simplified with respect to the case of conventional \ac{LDPC} codes. 
This also makes the new family of codes suitable for real-time communication systems, such as video conferencing or online gaming, where low latency is a critical requirement. The small encoding/decoding complexity of \ac{RC-LDPC} codes makes them suitable also for energy-efficient communications systems, such as \ac{IoT} devices, which are characterized by low power and computational capacity. 
For all these reasons, \ac{RC-LDPC} codes are actually employed in 5G \cite{Hui2018} and will likely remain a critical component of future communication systems and standards.

\subsection{Our contribution}

The goal of this paper is threefold. First and foremost, we propose a new family of \ac{RC-LDPC} codes, called \ac{PRC-LDPC} codes.  Secondly, we give new insights on how Golomb rulers can be employed to design \ac{LDPC} codes. Finally, we demonstrate the possibility of utilizing very simple encoders for codes in this family that exhibit comparable  performance with respect to the best codes available in the literature.

Our design starts from simplex codes, but we show that the primitive polynomial representing the parity-check matrix of the code needs to comply with some additional constraints in order to be an \ac{LDPC} code. Namely, first of all, we show that to make the code able to satisfy the \ac{RCC} \cite{RyanBook}, the support of the coefficients vector of the primitive polynomial must be a Golomb ruler. We also show that this is a necessary but not sufficient conditions to obtain a good \ac{LDPC} code. In fact, inaccurate choices of the polynomial can yield poor Hamming weight distributions and very poor minimum distance properties for the code. Therefore, we provide several rules to avoid these undesirable occurrences. In order to obtain a very fine rate-compatibility, we make use of puncturing \cite{ElKhamy2009}, in such a way that both high code rates and low code rates can be achieved. If needed, also shortening operations can be performed. The minimum distance profile of these codes is then analyzed. Many theoretical results are provided, and a method to estimate the minimum distance of these codes is discussed. Numerical examples show that the proposed method yields a good predictability of minimum distance properties. We finally show that the \ac{PRC-LDPC} codes can be encoded by using a very simple encoder circuit, and have a good error rate performance even when they are decoded with the low-complexity versions of \ac{BP}  algorithms typical of \ac{LDPC} codes.

\subsection{Related works}

The literature contains a plethora of works on rate compatible codes and codes whose design is based on Golomb rulers. In this section we briefly describe those most related to our proposal.

Some methods utilize modular Golomb rulers to construct circulant matrices that form the parity-check matrix of binary \ac{QC-LDPC} codes, as described in \cite{Ivanov2017,Xiao2021,Kim2022}. Similarly, non-binary LDPC codes that employ modular Golomb rulers are developed in \cite{Chen2012,Zhao2016}. Compared with the other methods, \ac{PRC-LDPC} codes are: (i) binary, (ii) non-QC, and (iii) based on standard (non-modular) Golomb rulers. The latter issue is particularly important as it is indeed the removal of the modular constraint that leads to the bit-level rate adaptivity of the new family of codes.

In \cite{Shirvanimoghaddam}, primitive rateless codes are designed. It is recognized that these codes can be realized as punctured simplex codes. However, differently from our codes, primitive rateless codes are not necessarily \ac{LDPC} codes and are not treated as such. The inclusion of the requirement on the sparsity of the parity-check matrix deserves a self-standing analysis, taking into account the peculiarities of \ac{LDPC} codes and their decoding algorithms.  In the same paper, many results on the average Hamming weight distribution of these codes are provided.
We instead do not restrict ourselves to studying the average case, but analyze the minimum distance properties for any individual code, based on its parity-check polynomial. Furthermore, we show that the satisfaction of the \ac{RCC}, necessary for \ac{LDPC} codes, leads to new results on the code minimum distance, which do not necessarily hold if length-$4$ cycles are present in the code Tanner graph \cite{Tanner1981}, the latter actually being the setting of most previous works, where punctured simplex codes are not considered as \ac{LDPC} codes.

In \cite{BaldiSimplex} punctured simplex codes are employed for error detection rather than error correction. Also in this case, the designed codes are not \ac{LDPC} codes. The provided results on the weight distribution hold for values of the block length $n>2^{k-1}$.
We instead employ values of $n$ that in most cases do not satisfy the above condition, and therefore do not (and cannot) make use of the above results. 

This work is also related to its preliminary shorter versions \cite{Battaglioni2022Fitce,Battaglioni2023ICC}. In those papers, the authors introduced the problem and provided a preliminary theoretical analysis, but many results are partial, or valid only for specific values of the code rate. Moreover, only loose empirical estimates of the minimum distance are provided, which may not hold for all the considered codes. In this paper we generalize, enrich and make both the theoretical and the numerical treatment more robust, by providing a comprehensive outline of the proposed family of codes. 

\subsection{Outline of the paper}

The paper is organized as follows. In Section \ref{sec:preli} we introduce the notation and recall the necessary background. In Section \ref{sec:sequense} we describe our design and discuss its requirements. In Section \ref{sec:rules} we give some rules for the choice of the primitive polynomial and provide theoretical results and examples. In Section \ref{sec:complex} we discuss encoding and decoding complexity. In Section \ref{sec:montecarlo} we assess the error rate performance of the newly designed codes. Section \ref{sec:conclusions} concludes the paper.

\section{Preliminaries}\label{sec:preli}



We use the notation $[a,b]$ to represent the set of integers between $a$ and $b$, including the endpoints.  A sequence of non-negative integers where every difference between two integers is distinct is called a Golomb ruler. The Hamming weight of a vector is the number of non-zero symbols it contains and in the following we simply refer to it as its weight. Similarly, the weight of a polynomial is the number of its non-zero coefficients. To every polynomial $h(x) = h_0 + h_1 x + ... + h_k x^k$, we associate a coefficients vector $\6h = (h_0, ..., h_k) $. The reciprocal of a polynomial $h(x)$ is denoted as $h^*(x)=x^kh(x^{-1})$. Since it is widely employed, the Hamming weight of $\6h$ is denoted as $w_h$. A polynomial of degree $k$ in $\mathbb{F}_2[x]$ is said to be primitive if it is the minimal polynomial of a primitive element of $\mathbb{F}_{2^k}$. It is well-known that primitive polynomials have an odd weight.

Given the finite field $\mathbb F_q$ with order $q$, a code $C$ is a $k-$dimensional subspace of $\mathbb F_q^n$, where $k<n$. The codewords in $C$ can be obtained as $C = \{ \mathbf c\in\mathbb F_q^n | \mathbf c \mathbf H^\top = \mathbf 0 \}$, where $^\top$ denotes transposition, and $\mathbf H\in\mathbb F_q^{r\times n}$ is a full-rank matrix of size $r \times n$, where $r = n-k$, and is known as the parity-check matrix. The code rate $R$ is defined as $R=\frac{k}{n}$. The number of codewords of weight $w$ is denoted as $A(w)$. The minimum distance of the code, denoted as $d$, is the smallest positive value of $w$ such that $A(w)>0$.

\ac{LDPC} codes are a special type of code characterized by parity-check matrices with a relatively small number of non-zero entries compared to the number of zeros. The \ac{RCC}  in the parity-check matrix specifies that no closed length-$4$ cycle is formed in the corresponding Tanner graph\cite{Tanner1981}, i.e., the parity-check matrix has no groups of four non-zero entries at the vertices of a rectangle. It is well known that soft-decision decoding algorithms commonly used for \ac{LDPC} codes, such as the sum-product algorithm\cite{Kschischang}, encounter convergence issues when the parity-check matrix contains the aforementioned length-$4$ cycles. For the rest of this paper, we assume that $q=2$.

A linear block code is said to be cyclic if the cyclic-shift of each of its codeword is also a codeword. To obtain the generator polynomial $g(x)$ of a cyclic code from a parity-check polynomial 
\[
h(x)=h_0+h_1x+\ldots+h_kx^k,
\]
that is a factor of $x^N+1$ in $\mathbb{F}_2[x]$, it is sufficient to divide $x^N+1$ by $h^*(x)$, which is also a factor of $x^N+1$ \cite{RyanBook}. The resulting cyclic code has non-zero codewords that can be expressed as\cite{Golomb}
\[
t(x)=x^jg(x)\mod{x^N+1}, \hspace{1em} j\in[0,N-1].
\] 
If $h(x)$ is a primitive polynomial, the code obtained by taking it as the parity-check polynomial is a cyclic simplex code with a block length of $N=2^k-1$. Therefore, this simplex code is formed by the all-zero codeword and all the cyclic shifts of any non-zero codeword. The following important property holds.
\begin{Proper}
Given any non-zero $k$-tuple $\6z$ and a non-zero codeword $\6t$ of the cyclic simplex code, there is exactly one set of $k$ consecutive entries of $\6t$ (including those wrapping around) which is equal to $\6z$.  
\label{pro:progword}
\end{Proper}
An important consequence of Property \ref{pro:progword} is the following one.

\begin{Proper}\cite{Fredricsson1975}
Any non-zero codeword of the cyclic simplex code is a pseudo-noise sequence of maximum period $N=2^k-1$, since $h(x)$ is primitive. 
\label{pro:prognoise}
\end{Proper}

We point out that, in the rest of the paper, when looking for specific $k$-tuples in the pseudo-noise sequence, the latter always ``wraps around'', i.e., its first entry and its last entry  are considered consecutive. We denote the pseudo-noise sequence associated to the codeword(s) of the cyclic simplex code as $\6p$.

The parity-check matrix $\mathbf{H}$ of the simplex code has $2^k-1-k$ rows, where $\6h$, i.e., the coefficients vector of $h(x)$, shifts from left to right by one position for each row. We define the support of $\6h$ as a vector $\mathbf{e}$ with $w_h$ elements. The vector $\mathbf{e}$ contains all $\{i \in [0,k] | h_i=1 \}$ in ascending order. We also define a vector $\mathbf{s}$ with $w_h-1$ elements, where $s_i=e_{i+1}-e_{i}$, $i\in \{0,\ldots,w_h-2\}$. It holds that \[
\sum_{i=0}^{w_h-2}s_i=k.
\] 
In the following, we will call \textit{separations} the entries of $\6s$ and distinguish between \textit{external} separations ($s_0$ and $s_{w_h-2}$) and \textit{internal} separations ($s_i$ with $i\in[1,w_h-3]$).
It is easy to observe that the maximum number of distinct separations between (not necessarily consecutive) pairs of non-zero entries in $\mathbf{h}$ is $\binom{w_h}{2}$. 
Finally, the largest entry of $\6s$ is denoted as $s_{\max}$.

\section{Design principles of PRC-LDPC codes}
\label{sec:sequense}
In this Section we describe the fundamental requirements of our code design.

\subsection{Parity-check matrix of PRC-LDPC codes}

According to the description given in Section \ref{sec:preli}, the form of the parity-check matrix of \ac{PRC-LDPC} codes is shown in Fig. \ref{fig:hriferimento}.

\begin{figure}[thb]
\centering
\begin{equation*}
\small \mathbf{H} =
\begin{tikzpicture}[baseline=(current bounding box.center)]
 \matrix[matrix of math nodes,left delimiter=(,right delimiter=),scale=0.7] (m)
   {
h_0 & h_1 & \cdots & h_k & & &&&& \\
& h_0 & h_1&\cdots  & h_k &&&&& \\
& & \ddots & \ddots & \vdots &&& \\
& & & h_0 & h_1 & \cdots & h_k &&&\\
& & & & h_0 & h_1 & \cdots & h_k &&\\
& & & & & h_0 & h_1 & \cdots & h_k& \\
};
\end{tikzpicture}
\end{equation*}
\caption{General form of the parity-check matrix of the codes considered in this paper.}
\label{fig:hriferimento}
\end{figure}

From the parity-check matrix of the original simplex code, for which $n=N=2^k-1$ and therefore $R=\frac{k}{2^k-1}$, puncturing can be used to obtain the parity-check matrix of a code with larger rate. 
An example is shown in Fig. \ref{fig:H1H2}, where the last two rows have been eliminated. After performing a certain number of puncturing operations, the parity-check matrix is reduced to $r$ rows and $n=r+k$ columns.  Referring to Fig. \ref{fig:H1H2}, we notice that the non-zero codewords of the punctured code, whose number is $2^k-1$ as in the original simplex code, can be obtained as the vectors selected by a sliding window of length $n<2^k-1$ that circularly spans over the pseudo-noise sequence introduced in Property \ref{pro:prognoise}, i.e., circularly spans over a non-zero codeword of the parent simplex code. This unique origin of the codewords enables us to investigate the generation of low-weight codewords. 

\begin{figure}[thb]
\centering
\begin{equation*}
\small \mathbf{H} =
\begin{tikzpicture}[baseline=(current bounding box.center)]
 \matrix[matrix of math nodes,left delimiter=(,right delimiter=),scale=0.7] (m)
   {
h_0 & h_1 & \cdots & h_k & & &&&& \\
& h_0 & h_1& \cdots  & h_k &&&&& \\
& & \ddots & \ddots & \vdots &&& \\
& & & h_0 & h_1 & \cdots & h_k &&&\\
& & & & h_0 & h_1 & \cdots & h_k &&\\
& & & & & h_0 & h_1 & \cdots & h_k& \\
};
\draw[red] (-3,-1) -- (3,-1);
\draw[red] (-3,-1.5) -- (3,-1.5);
\draw[red] (1.9,-1.8) -- (1.9,1.8);
\draw[red] (2.5,-1.8) -- (2.5,1.8);
\end{tikzpicture}
\end{equation*}
\caption{Effect of puncturing on the parity-check matrix.}
\label{fig:H1H2}
\end{figure}

Let us discuss the requirements that $h(x)$ must fulfill in order to define a \ac{PRC-LDPC} code.

\subsection{Requirements for the parity-check polynomial}


First of all, since the design starts from simplex codes, $h(x)$ is a primitive polynomial.  Differently from previous works, in this paper we need to tighten the requirements on $h(x)$. The following result holds.

\begin{The}
A necessary and sufficient condition for the satisfaction of the \ac{RCC} in a \ac{PRC-LDPC} code is that $\6e$, i.e., the support vector of the coefficients vector of $h(x)$, is a Golomb ruler.
\label{the:theGol}
\end{The}
\begin{IEEEproof}
A length-$4$ 
cycle exists in $\mathbf{H}$ if and only if there exist two pairs $(i_1,i_2)$, $(i_3,i_4)$ such that $h_{i_1}=h_{i_2}=h_{i_3}=h_{i_4}=1$ and $i_2-i_1=i_4-i_3$, being $(i_1,i_2,i_3,i_4)$ different one another, except that it might be $i_1=i_4$.

Each entry of $\mathbf{e}$ is associated to a non-zero coefficient of $h(x)$. Then, if $\mathbf{e}$ is a Golomb ruler, by definition, there cannot exist two pairs of different indices $(j_1,j_2)$ and $(j_3,j_4)$ such that $e_{j_2}-e_{j_1}=e_{j_4}-e_{j_3}$. However, if $\mathbf{e}$ contains $e_k$, for some $k$, then, by definition, $h_{e_k}=1$. Therefore, if $\mathbf{e}$ is a Golomb ruler, there cannot exist two pairs $(i_1,i_2)$, $(i_3,i_4)$ such that $h_{i_1}=h_{i_2}=h_{i_3}=h_{i_4}=1$ and $i_2-i_1=i_4-i_3$. This implies that, if $\mathbf{e}$ is a Golomb ruler, $\mathbf{H}$ cannot contain length-$4$ cycles and therefore satisfies the row-column constraint.

In order to prove that this condition is also necessary we need to show that, if $\mathbf{e}$ is not a Golomb ruler, then $\mathbf{H}$ does not satisfy the row-column constraint. If $\mathbf{e}$ is not a Golomb ruler, then there exist two pairs $(j_1,j_2)$ and $(j_3,j_4)$ such that $e_{j_2}-e_{j_1}=e_{j_4}-e_{j_3}$, also implying that $h_{e_{j_1}}=h_{e_{j_2}}=h_{e_{j_3}}=h_{e_{j_4}}=1$. This is the condition of existence of a length-$4$ cycle, which corresponds to invalidity of the \ac{RCC}.
\end{IEEEproof}

Therefore, in the following we choose the parity-check polynomial in such a way that Theorem \ref{the:theGol} holds. This way, a fundamental condition required for effective convergence of the iterative decoders commonly used to decode LDPC codes is satisfied.

At this point, we also need to make some considerations on the sparsity of the code and of the primitive parity-check polynomial. Parity-check matrices as in Fig. \ref{fig:hriferimento} are row-regular, since all the rows have Hamming weight equal to $w_h$, but are not column-regular. The average column weight is \begin{equation}
    \langle w_c \rangle =\frac{rw_h}{n}=(1-R)w_h.
    \label{eq:wcmedio}
\end{equation}
This parameter plays a crucial role in terms of iterative decoding complexity, which increases for increasing values of $\langle w_c \rangle$. 
Therefore, in order to keep a low decoding complexity, for a given value of $R$, we will choose a relatively small value of $w_h$. Complexity issues will be discussed more in depth in Section \ref{sec:complex}.

We remark that having a small $w_h$ may not be sufficient to guarantee that $\6e$ is a Golomb ruler. For example, suppose that $k=w_h=5$. Then, even though $\langle w_c \rangle$ is small, $h(x)=1+x+x^3+x^4+x^5$ and $\6e=[0,1,3,4,5]$ is not a Golomb ruler, leading to a lot of length-$4$ cycles and therefore a potentially bad performance under iterative decoding. We thus need to include the condition that $w_h$ is relatively small compared to $k$. In particular, a necessary, though not sufficient, condition for the fulfilment of the \ac{RCC} derives from the following statement.

\begin{Cor}
Given a parity-check polynomial $h(x)$, if $\binom{w_h}{2}>k$ then the \ac{RCC} does not hold for any \ac{PRC-LDPC} code described by $h(x)$. 
\label{cor:corsidon}
\end{Cor}
\begin{IEEEproof}
    As noted in Section \ref{sec:preli}, a polynomial $h(x)$ of weight $w_h$ yields $\binom{w_h}{2}$ positive separations between its exponents. In order to satisfy the \ac{RCC}, due to Theorem \ref{the:theGol}, we need all of them to be different. If the degree of $h(x)$ is $k$, then the separations must take values in $[1,k]$ (which obviously contains $k$ elements). Therefore, the pigeonhole principle tells us that if $\binom{w_h}{2}>k$, i.e., if there are more separations than available different values, then at least two separations are equal and, therefore, the \ac{RCC} is not satisfied.
\end{IEEEproof}

We would like to remark that, in order to satisfy all the needed constraints, i.e., to find a polynomial that is both primitive, sparse and such that $\6e$ is a Golomb ruler, we may need values of $w_h$ that are not necessarily tight to the bound given by Corollary \ref{cor:corsidon}. 

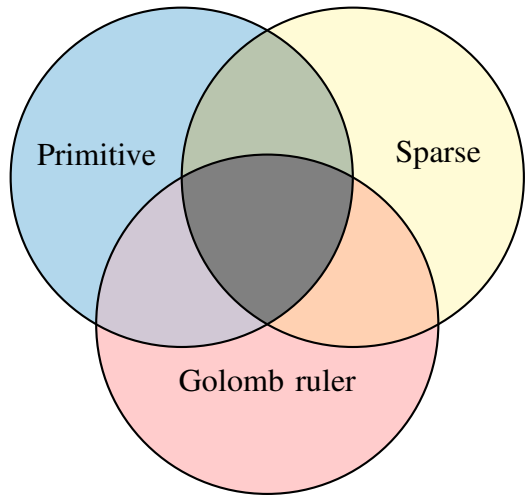
\begin{figure}[thb]
    \centering
    \begin{tikzpicture}[scale=1.5]
  \def\radius{1.5};
  \def\cOne{(0,0)};
  \def\cTwo{(60:\radius)};
  \def\cThree{(120:\radius)};

    \fill[myblue2!30] \cThree circle (\radius);
    \fill[red!20] \cOne circle (\radius);
    \fill[yellow!20] \cTwo circle (\radius);

 \begin{scope}
    \clip \cOne circle (\radius);
    \clip \cTwo circle (\radius);
    \fill[red!60!yellow!30] \cOne circle (\radius);
  \end{scope}
   \begin{scope}
    \clip \cOne circle (\radius);
    \clip \cThree circle (\radius);
    \fill[myblue2!60!red!30] \cOne circle (\radius);
  \end{scope}
   \begin{scope}
    \clip \cTwo circle (\radius);
    \clip \cThree circle (\radius);
    \fill[yellow!50!myblue2!60] \cTwo circle (\radius);
  \end{scope}
  
  \begin{scope}
    \clip \cOne circle (\radius);
    \clip \cTwo circle (\radius);
    \clip \cThree circle (\radius);
    \fill[gray, even odd rule] \cOne circle (\radius) \cTwo circle (\radius) \cThree circle (\radius);
  \end{scope}
  \draw[thick] \cOne circle (\radius);
  \draw[thick] \cTwo circle (\radius);
  \draw[thick] \cThree circle (\radius);
  \node at (0,-0.5) {Golomb ruler};
  \node at (1.5,1.5) {Sparse};
  \node at (-1.5,1.5) {Primitive};
\end{tikzpicture}
    \caption{Requirements of the parity-check polynomial.}
    \label{fig:Venn}
\end{figure}

A visual summary of the concepts discussed in this section is depicted in Fig. \ref{fig:Venn}, where the searched polynomials live in the gray region. 

\subsection{Shortening operations: a further degree of freedom}

Aiming to gain an even finer rate adaptability, it is possible to apply code shortening. This operation consists in eliminating $z$ information symbols. Therefore, the shortened code has block length $n'=n-z$ and $k'=k-z$ information symbols. The parity-check matrix of the shortened code can be obtained from $\mathbf{H}$ by deleting the first, or the last, $z$ columns (or even a combination of them). A simple example is shown in Fig. \ref{fig:H1H2s}, where the first two columns have been eliminated. 

\begin{figure}[thb]
\centering
\begin{equation*}
\small \mathbf{H} =
\begin{tikzpicture}[baseline=(current bounding box.center)]
 \matrix[matrix of math nodes,left delimiter=(,right delimiter=),scale=0.7] (m)
   {
h_0 & h_1 & \cdots & h_k & & &&&& \\
& h_0 & h_1& \cdots  & h_k &&&&& \\
& & \ddots & \ddots & \vdots &&& \\
& & & h_0 & h_1 & \cdots & h_k &&&\\
& & & & h_0 & h_1 & \cdots & h_k &&\\
& & & & & h_0 & h_1 & \cdots & h_k& \\
};

\draw[red] (-1.9,-1.8) -- (-1.9,1.8);
\draw[red] (-2.5,-1.8) -- (-2.5,1.8);
\end{tikzpicture}
\end{equation*}
\caption{Effect of shortening on the parity-check matrix.}
\label{fig:H1H2s}
\end{figure}

\section{Code design approach}
\label{sec:rules}
We have shown in Section \ref{sec:sequense} that the parity-check polynomial must comply with certain basic rules. In this section we provide sufficient conditions for the existence of low-weight codewords in punctured simplex codes and convert them into a practical approach to code design based on good practice.

\subsection{Low-weight codewords}\label{subsec:lowen}

We have mentioned in Section \ref{sec:preli} that any non-zero codeword of the parent cyclic simplex code corresponds to a shifted version of the same pseudo-noise sequence $\6p$ of length $N=2^k-1$. 

\begin{Rem}
When puncturing is applied, and a code of length $n<N$ is obtained, the code is no longer cyclic and each codeword is a different $n$-tuple of consecutive symbols of $\6p$. In other words, it is possible to consider a sliding window of size $n$ spanning over $\6p$. Each of the possible $N$ initial positions of the sliding window covers a different non-zero codeword.
\label{rem:sliwi}
\end{Rem}

Remark \ref{rem:sliwi} is fundamental for our theoretical analysis and its implications will be used many times in the following. Therefore, we provide a toy example to make it clearer.

\begin{Exa}
Consider the $(7,3,4)$ cyclic symplex code characterized by the parity-check polynomial $h(x)=1+x^2+x^3$. The corresponding pseudo-noise sequence is given by $g(x)=1+x+x^2+x^4$ followed by $N-r-1=2$ zeros, i.e.,
\[
\6p=[1,1,1,0,1,0,0].
\]
Let us suppose that we puncture two code symbols, obtaining a $(5,3)$ code. The $2^k-1=7$ non-zero codewords of this code can be obtained by letting a window of size $5$ cyclically slide over $\6p$, for all possible initial positions, that is,
\begin{align*}
[\mathbf{1,1,1,0,1},\color{black!30}0,0\color{black}],\\
[\color{black!30} 1 \color{black},\mathbf{1,1,0,1,0},\color{black!30}0\color{black}],  \\ 
[\color{black!30}1,1,\color{black}\mathbf{1,0,1,0,0}],\\
[ \mathbf{1}, \color{black!30}1,1,\color{black}\mathbf{0,1,0,0}],  \\ 
[ \mathbf{1,1}, \color{black!30}1,0,\color{black}\mathbf{1,0,0}],  \\ 
[ \mathbf{1,1,1}, \color{black!30}0,1,\color{black}\mathbf{0,0}],  \\ 
[ \mathbf{1,1,1,0}, \color{black!30}1,0,\color{black}\mathbf{0}],  \\ 
\end{align*}
where the codewords of the punctured code are marked in bold.

\end{Exa}

Keeping these concepts in mind, we are interested in finding sub-sequences of $\6p$ that may cause the existence of low-weight codewords. We first focus on the following two tuples, called $\mathbf{T}_1$ and $\mathbf{T}_2$ in the following, respectively: 
\begin{enumerate}
    \item a $k$-tuple containing $k-1$ zeros and a single one;
    \item a $(k+1)$-tuple coinciding with $\6h^*$.
\end{enumerate}
Whereas the existence of $\mathbf{T}_1$ in $\6p$ is implied by Property \ref{pro:progword}, next we prove a sufficient condition for the existence of $\mathbf{T}_2$ in $\6p$.

\begin{The}
Given a primitive parity-check polynomial $h(x)$ defining a cyclic simplex code of length $N$, if $\6e$ is a Golomb ruler, then $\6p$ contains a $(k+1)$-tuple coinciding with $\6h^*$.
\label{the:exhstar}
\end{The}
\begin{IEEEproof}
    Let us evaluate the quantity $\6h^*\6h^\top \mod 2$. If $k$ is odd we have that 
    \begin{equation}
       \6h^*\6h^\top \mod 2=\begin{bmatrix}
h_k & h_{k-1} & \cdots & h_1& h_0 
\end{bmatrix}
\begin{bmatrix}
h_0 \\
h_1 \\
\vdots \\
h_{k-1}\\
h_k \\
\end{bmatrix} \mod 2
= \bigoplus_{i=0}^{k}h_ih_{k-i}=0, 
\label{eq:oddcase}
    \end{equation}

 since the sum contains an even number of pairs of equal terms.

If $k$ is even, instead we have that $\6h^*\6h^\top$ is
 \begin{equation}
\bigoplus_{i=0}^{k}h_ih_{k-i}
= h_0h_k + \cdots +  h_{k/2}h_{k/2} +  \cdots + h_kh_0,
\label{eq:evencase}
    \end{equation}
which is $0$ modulo $2$ only if $h_{k/2}=0$. If $\6e$ is a Golomb ruler, $h_{k/2}$ is necessarily $0$. In fact, $h_{k/2}=1$ implies that $h_0$ and $h_{k/2}$ are at the same distance of $h_k$ and $h_{k/2}$, which contravenes the definition of Golomb ruler for $\6e$. In summary, if $\6e$ is a Golomb ruler, we have that  $\6h^*\6h^\top = 0 \mod 2$.

Given any codeword  $\6t$ of the cyclic simplex code, since the rows of the parity-check matrix are formed by cyclic shifts of the vector $[\6h,\60_{N-k-1}]$, from $\6t\6H^\top=\60$ it follows that
\begin{equation}
\bigoplus_{i=0}^{k} t_{i} h_{i}=0. 
\end{equation}
From the comparison with \eqref{eq:oddcase} and \eqref{eq:evencase} we notice that, if $\6e$ is a Golomb ruler, we can substitute $t_i=h^*_i$, for $i\in[0,k]$. This means that any codeword of the cyclic simplex code contains $\6h^*$ and, as codewords are cyclically shifted versions of $\6p$, also $\6p$ contains $\6h^*$, proving the thesis. 
\end{IEEEproof}

We now show that, under certain circumstances, the relative position of $\mathbf{T}_1$ and $\mathbf{T}_2$ can be predicted.

\begin{Lem}
If, for a family of \ac{PRC-LDPC} codes,
$$
 s_0>\sum_{j\in[1,w_h-2]}s_j, \hspace{2em} \mathrm{or} \hspace{2em}  s_{w_h-2}>\sum_{j\in[0,w_h-3]}s_j,
$$
and $\6e$ is a Golomb ruler, then $\mathbf{T}_1$ and $\mathbf{T}_2$ are consecutive in $\6{p}$.
\label{lem:lemmaZ}
\end{Lem}
\begin{IEEEproof}
By definition of simplex codes, $\6p$ is given by 
\[\6p=[\6g , \60_{k-1}],\]
where  $\6g$ is the coefficients vector of the generator polynomial, and $\60_{k-1}$ is a $(k-1)$-tuple of zeros.
Thus, since $g_0=g_{n-k-1}=1$, in order to prove the thesis we just need to show that, under the above hypotheses, either $\6g$ begins or ends with $\6h^*$ (i.e.,  $\mathbf{T}_2$), which is for sure contained in $\6p$ due to Theorem \ref{the:exhstar}. Let us suppose that \begin{equation}
s_{w_h-2}>\sum_{j\in[0,w_h-3]}s_j,
\label{eq:lastbig}
\end{equation}
that is, the first separation between the exponents of $h(x)$ is larger than the sum of all the others. This implies that the last separation between the exponents of $h^*(x)$ is larger than the sum of all the others. 
To study the structure of $\6g$ we can perform classical polynomial division (from left to right) between $x^N+1$ and $h^*(x)$. The first partial remainder of the division is $$r_0(x)=x^{N-k}h^*(x)+x^N=\sum_{i=0}^{w_h-2}x^{N-\sum_{j=0}^is_j}$$ (in $\mathbb{F}_2[x]$), and the first  quotient is obviously $q_0(x)=x^{N-k}$. The largest and the smallest exponent of $r_0(x)$ are therefore $x^{N-s_0}$ and $x^{N-k}$, respectively. Then, we notice that the next term of the quotient must be $x^{N-k-s_0}$ and that the corresponding $h^*(x)x^{N-k-s_0}$ is in the form $x^{N-s_0}$ plus some terms with degree lower than $x^{N-k}$, due to \eqref{eq:lastbig}. The same reasoning holds for the remaining $w_h-2$ terms of the quotient. Summarizing, the first $w_h$ terms of the quotient (that is $g(x)$) are
\[x^{N-k}, x^{N-k-s_0}, x^{N-k-(s_0+s_1)}, \ldots, x^{N-2k}, \]
that is a shifted version of $h^*(x)$ (in particular, it is $x^{N-2k}h^*(x)$). Therefore, the first $k+1$ bits of $\6g$ are $\6h^*$, implying that $\mathbf{T}_2$ follows $\mathbf{T}_1$ in $\6p$.

The specular case, in which $$s_{0}>\sum_{j\in[1,w_h-2]}s_j,$$ is perfectly equivalent, but it is convenient to perform the polynomial division from right to left. In this case, the last $k+1$ symbols of $\6g$ are $\6h^*$ and $\mathbf{T}_2$ precedes $\mathbf{T}_1$ in $\6p$.
\end{IEEEproof}

Lemma \ref{lem:lemmaZ} has important implications on the minimum distance of some codes.

\begin{Cor}
Given any \ac{PRC-LDPC} code of rate $R\geq \frac{1}{2}$, if $$s_0>\sum_{j\in[1,w_h-2]}s_j, \hspace{2em} \mathrm{or} \hspace{2em}  s_{w_h-2}>\sum_{j\in[0,w_h-3]}s_j,$$
and satisfying the \ac{RCC}, then $d\leq w_h$.
\end{Cor}
\begin{IEEEproof}
    It follows from Lemma \ref{lem:lemmaZ} that, under its hypotheses, $\6p$ contains a $2k$-tuple of weight $w_h$, formed by the concatenation of the $k-1$ zeros in  $\mathbf{T}_1$ and $\mathbf{T}_2$ (of size $k+1$ and weight $w_h$). Therefore, since the codewords of punctured simplex codes can be covered by a window of size $n$ sliding over $\6p$, all the punctured simplex codes of block length $n\leq 2k$ satisfying the hypotheses of Lemma  \ref{lem:lemmaZ} contain at least one codeword of weight smaller than or equal to $w_h$.
\end{IEEEproof}

In order to better understand the role of large external separations, let us deepen the analysis for the case of $R=\frac{1}{2}$, and define the following sufficient condition for the existence of codewords of weight $w_h$ when an external separation is larger than the sum of the internal ones.

\begin{Lem}
In a \ac{PRC-LDPC} code with $R=\frac{1}{2}$ satisfying the \ac{RCC}, if for, either $i=0$ or $i=w_h-2$,
\[
s_i>\sum_{j\in[1,w_h-3]}s_j,
\] 
then there are at least
$$s_i-\sum_{j\in[1,w_h-3]}s_j$$
codewords of weight $w_h$.
\label{lem:whprimo}
\end{Lem}
\begin{IEEEproof}
The proof can be found in Appendix \ref{app:appendiceprova}.
\end{IEEEproof}

Following the above reasoning, we can also analyze the case in which $s_0+s_{w_h-2}$ exceeds the sum of the internal separations, and also the case in which an internal separation exceeds the sum of all the others. These two conditions cannot be simultaneously true. The following two lemmas can be proved using similar arguments as those used in the proof of Lemma \ref{lem:whprimo}, and their proofs are hence omitted for brevity.

\begin{Lem}
In a \ac{PRC-LDPC} code with $R=\frac{1}{2}$ satisfying the \ac{RCC},  if 
\begin{equation}
    s_0+s_{w_h-2}>\sum_{j\in[1,w_h-3]}s_j,
 \label{cond1s} 
\end{equation} 
then there are at least $$s_0+s_{w_h-2}-\sum_{j\in[1,w_h-3]}s_j$$ codewords of weight $w_h$.

\label{lem:whsecondo}
\end{Lem}

\begin{Lem}
In a \ac{PRC-LDPC} code with $R=\frac{1}{2}$  satisfying the \ac{RCC}, if there exists $i\in[1,w_h-3]$ such that $$s_i>\sum_{j\neq i}s_j,
$$ 
then there are  at least
\begin{equation}
    s_i-\sum_{i\neq j}s_j
    \label{conds2}
\end{equation}
codewords of weight $w_h$.
\label{lem:whterzo}
\end{Lem}

It should be noted that Lemmas \ref{lem:whprimo}. \ref{lem:whsecondo} and \ref{lem:whterzo} remain applicable even for $R>\frac{1}{2}$. Specifically, if a particular code satisfies $d>w_h$ at $R=\frac{1}{2}$ and is subsequently punctured to attain $d=w_h$, then the methodologies outlined in Lemmas \ref{lem:whsecondo} and \ref{lem:whterzo} can be utilized to determine the multiplicity of low-weight codewords.

\begin{figure}[thb]
\centering
\begin{tikzpicture}
    \draw (0,0)  rectangle (14,11);
    \draw (7,7.5) [fill=gray!10] ellipse (6 and 3);
    \draw (7,2) [fill=gray!10]  ellipse (4 and 1.5);
     \draw (7,0.5) -- (7,3.5);
    \node at (7,5.25) {$s_0+s_3>s_1+s_2$};
    \node at (5,2) {$s_1>s_0+s_2+s_3$};
    \node at (9,2) {$s_2>s_0+s_1+s_3$};

 \draw (9.5,7.7) ellipse (3 and 1.7);
    \draw (4.5,7.7) ellipse (3 and 1.7);
\node at (4.4,8.8) [font=\small]{$s_0>s_1+s_2$};
\node at (9.6,8.8) [font=\small] {$s_3>s_1+s_2$};

\draw (10.45,7.5) ellipse (1.4 and 0.8);
    \draw (3.55,7.5) ellipse (1.4 and 0.8);
        \node  at (10.5,7.5)  [font=\scriptsize]{$s_3>s_0+s_1+s_2$};
    \node at (3.55,7.5) [font=\scriptsize]{$s_0>s_1+s_2+s_3$};
\end{tikzpicture}
\caption{Summary of conditions causing $d\leq w_h$, when $w_h=5$.}
\label{fig:ellissi}
\end{figure}
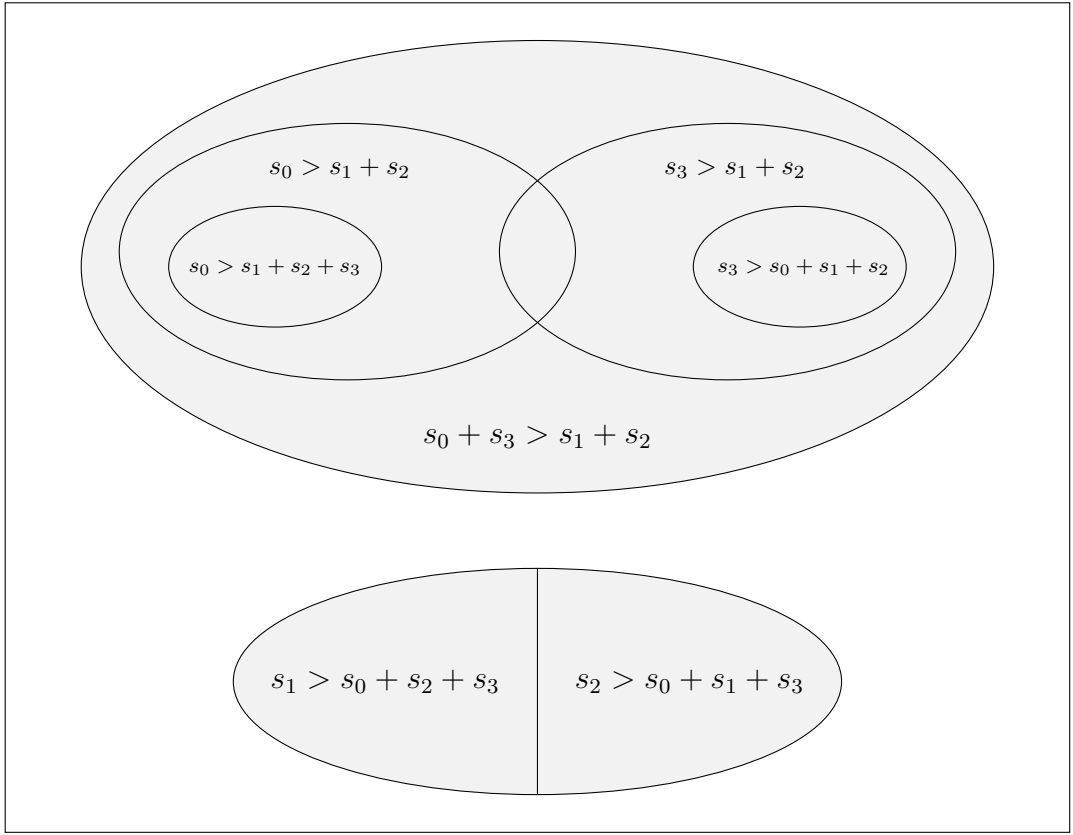

We remark that the hypotheses of Lemmas \ref{lem:lemmaZ} to \ref{lem:whterzo} may imply each other. For example, the hypotheses of Lemma \ref{lem:lemmaZ} imply those of Lemma \ref{lem:whprimo}. 
In Fig. \ref{fig:ellissi} we provide an illustrative scheme for the cases in which $d\leq w_h=5$, and thus $|\6s|=4$. The gray region contains all the sufficient conditions causing the minimum distance to be upper bounded by $w_h$.

Summarizing, some further good practices, besides the strict rules provided in Section \ref{sec:sequense}, are given in the following remark.

\begin{Rem}
    When designing \ac{PRC-LDPC} codes, the parity-check polynomial should be chosen such that:
    \begin{itemize}
        \item the first and the last entries of $\6s$ are not too large (i.e., they do not exceed the sum of all the other separations, or the sum of the internal separations);
        \item any internal entry of $\6s$ is not too large (i.e., it does not exceed the sum of the other  separations);
        \item the sum of the first and the last entry of $\6s$ is not too large (i.e., it does not exceed the sum of the internal separations).
    \end{itemize}
    
    \label{rem:rul_ext_other}
\end{Rem} 

\begin{figure}[tbh!]
\centering
\begin{tikzpicture}
     \draw (-7,0) -- (7,0);
      \draw (-2,0) [fill=gray!80] rectangle (2,1);

\node at (0,1.7) {$\6h^*$};

\draw [decorate, decoration={brace, amplitude=8pt}]  (2, -0.1) -- (-2,-0.1); 

\node at (0,-0.7) {$k+1$};

\draw [decorate, decoration={brace, amplitude=8pt}]  (-2.05, -0.1) -- (-5,-0.1); 

\node at (-3.5,-0.7) {$Z_0$};

\draw [decorate, decoration={brace, amplitude=8pt}]  (5.5, -0.1) -- (2.05,-0.1); 

\node at (3.75,-0.7) {$Z_1$};

\draw[line width=2pt] (-5,0) -- (-5,1);
\draw[line width=2pt] (-5.6,0) -- (-5.6,1);
\draw[line width=2pt] (-6.8,0) -- (-6.8,1);

\node at (-6.2,0.5) {$\cdots$};

\draw [decorate, decoration={brace, amplitude=8pt}]  (-5.6,1.1) --(-5,1.1); 

\node at (-5.3,1.7) {$s_{0}$};

\draw[line width=2pt] (5.5,0) -- (5.5,1);
\draw[line width=2pt] (5.9,0) -- (5.9,1);
\draw[line width=2pt] (6.7,0) -- (6.7,1);

\node at (6.3,0.5) {$\cdots$};

\draw [decorate, decoration={brace, amplitude=8pt}]  (5.5,1.1) --(5.9,1.1); 

\node at (5.7,1.7) {$s_{w_h-2}$};
      
\end{tikzpicture}
\caption{Structure of $\6p$ around $\6h^*$.}
\label{fig:zonapettini}
\end{figure}
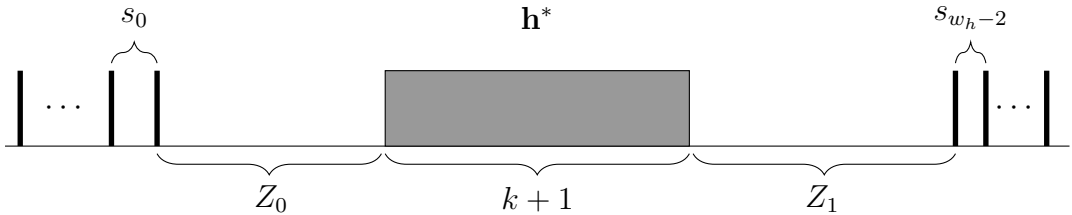

At this point, it is interesting to study the structure of $\6p$ when none of the sufficient conditions causing $d\leq w_h$ are satisfied. By computing $g(x)$ as the polynomial division of $x^N+1$ by $h^*(x)$, with the same arguments used in the proof of Lemma \ref{lem:lemmaZ}, it is straightforward to prove that the region around $\mathbf{T}_2$ appears as shown in Fig. \ref{fig:zonapettini}, if the \ac{RCC} is satisfied, along with some further requirements discussed in the following. Specifically, on either side of $\6h^*$, there exist two regions containing only zeros, of size $Z_0$ and $Z_1$, respectively. The size of these two regions is computed in the following theorem.

\begin{The}
    Given a \ac{PRC-LDPC} code characterized by a certain $h(x)$ of degree $k$, for which the conditions in Remark \ref{rem:rul_ext_other}  are satisfied, by defining $i^*$ as the index for which $s_{i^*}=s_{\max}$, we have that $$
Z_0=2\Big(\sum_{i=i^*}^{w_h-2}s_i\Big)-k-1,
$$
$$
Z_1= 2\Big(\sum_{i=0}^{i^*}s_i\Big)-k-1.
$$
\label{the:zerozones}
\end{The}
\begin{IEEEproof}
    The proof follows the same idea of the proof of Lemma \ref{lem:lemmaZ}, i.e., the regions around $\6h^*$ are studied by means of polynomial division between $x^{2^k-1}+1$ and $h^*(x)$. Therefore, for the sake of brevity, we omit it.
\end{IEEEproof}
 It easily follows from Theorem \ref{the:zerozones} that $$Z_0+Z_1=2(s_{\max}-1),$$
which has the very interesting implication that the total width of the all-zero region around $\6h^*$ depends only  on the largest separation between the exponents of $h(x)$.

Then, under the hypotheses of Theorem \ref{the:zerozones}, next to the all-zero region of size $Z_0$, there are some symbols one, each at distance $s_{w_h-2}$ from the other. Similarly, consecutive to the all-zero region of size $Z_1$, there are some symbols one, each at distance $s_0$ from the other. 

Let us now convert the above reasoning into conditions on the minimum distance of the analyzed codes.

\begin{Cor}
    Given any \ac{PRC-LDPC} code of block length $n\leq 2s_{\max}+k-1$ satisfying the hypotheses of Theorem \ref{the:zerozones}, then $d\leq w_h$.
    \label{cor:distanza_finestra}
\end{Cor}
\begin{IEEEproof}
    The proof follows from the 
    equality
    \[
    Z_0+Z_1+k+1=2\Big(\sum_{i=0}^{i^*}s_i+\sum_{i=i^*}^{w_h-2}s_i\Big)-k-1=2k+2s_{\max}-k-1=2s_{\max}+k-1.
    \]
    All the windows of size $n\leq Z_0+Z_1+k+1$ cover at least one codeword of weight smaller than or equal to $w_h$.
\end{IEEEproof}

Fig. \ref{fig:zonapettini} also provides a further confirmation that the external separations should not be too large. If this condition is met, sliding windows of size $n$ slightly larger than $Z_0+Z_1+k+1$ would cover codewords with relatively large weight.

\subsection{Families of codewords}\label{subsec:families}

\begin{figure}[thb]
\centering
\begin{tikzpicture}
\draw[black] (3.8,-0.25) rectangle (6.2,0.25);
 \node[circle,draw=black] at (0,0) {1};
  \node at (1,0) {0};
  \node at (2,0) {$\cdots$};
  \node at (3,0) {0};
   \node at (4,0) {1};
   \node at (5,0) {$\cdots$};
   \node at (6,0) {1};
  \node at (7,0) {0};
  \node at (8,0) {$\cdots$};
  \node at (9,0) {0};
  \node[circle,draw=black] at (10,0) {1};
  \draw [decorate, decoration={brace, amplitude=8pt}]  (3.2, -0.35) -- (0.8,-0.35); 
\node at (2,-0.95) {$Z_l$};
 \draw [decorate, decoration={brace, amplitude=8pt}]  (9.2, -0.35) -- (6.8,-0.35); 
\node at (8,-0.95) {$Z_r$};
 \draw [decorate, decoration={brace, amplitude=8pt}]  (6.25, -0.35) -- (3.75,-0.35); 
\node at (5,-0.95) {$L$};
\end{tikzpicture}
\caption{Portion of $\6p$ describing a family of codewords.}
\label{fig:family}
\end{figure}
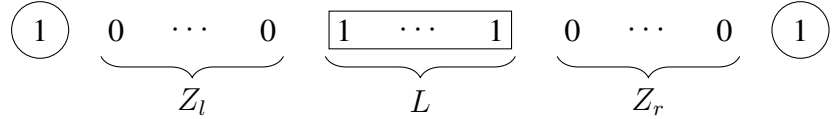

In this section we make some further considerations on the minimum distance of \ac{PRC-LDPC} codes.
For a given $n$,  we say that if two codewords can be obtained from each other by a non-cyclic shift, then they belong to the same \emph{family} of codewords. A family of codewords derives from a portion of $\6p$, as shown in Fig. \ref{fig:family}. Any window of size $n$ starting in the all-zero region of size $Z_l$ and ending in the all-zero region of size $Z_r$ covers codewords belonging to the same family. 

The family of codewords defined by the portion within the rectangle in Fig. \ref{fig:family}, say it is the $i$-th and is composed by codewords of weight $w$, exists for $n\in[L,L+Z_l+Z_r]$, i.e., until the window necessarily covers one of the two circled symbols one. It is interesting to count $A_i(w)$ for each value of $n$. By supposing, without loss of generality, that $Z_l\geq Z_r$, we have that
\begin{equation}
  A_i(w)=
\begin{cases}
n-L+1  & \text{if $L\leq n\leq L+Z_r$,} \\
1+Z_r & \text{if $L+Z_r<n\leq L+Z_l$,} \\
1+Z_r+Z_l+L-n & \text{if $L+Z_l<n\leq L+Z_r+Z_l$,} \\
0 & \text{elsewhere.}
\end{cases}
\label{eq:trapezio}
\end{equation}
Clearly, if $Z_r\geq Z_l$, we must swap 
 $Z_l$ and $Z_r$ in \eqref{eq:trapezio}.

We also introduce the quantities $r(d)$ and $n(d)$ which, given a parity-check polynomial $h(x)$, are the minimum number of rows and columns of $\6H$ such that the corresponding \ac{PRC-LDPC} code has distance $d$. To establish the starting point of our analysis, let us begin by examining an extreme scenario. While the codes described in Lemmas \ref{lem:lemr2} and \ref{lem:famiglia} may not be practical themselves, they play a valuable role in identifying regions within $\6p$ where low-weight codewords of codes with larger block length can potentially originate.

\begin{Lem}
A \ac{PRC-LDPC} code satisfying the \ac{RCC} has $d=1$ if and only if $r < r_{\min}=s_{\max}$. 
\label{lem:lemr2}
\end{Lem}
\begin{IEEEproof}
Let us suppose that the two entries of $\6h$ separated by $s_{\max}$ are $h_i$ and $h_j$. Since the \ac{RCC} is satisfied, there exists only one such pair. Then, there are $s_{\max}-1$ zeros separating $h_i$ and $h_j$. In order to avoid the occurrence of all-zero columns in the parity-check matrix, $r$ must necessarily be at least equal to $s_{\max}$. This way, all the columns from the $(i+1)$-th to the $(j-1)$-th contain at least one symbol one, that is $h_i$ shifting to the right for increasing values of $r$. If $r=\hat{r}<s_{\max}$, instead, due to the diagonal form of the parity-check matrix, $s_{\max}-\hat{r}$ columns are zero, i.e., those ranging from the $(i+\hat{r}-1)$-th to the $(j-1)$-th. The same reasoning can be applied to the other smaller separations. Therefore, if $r < s_{\max}$, the parity-check matrix contains columns of weight $0$, implying that $d=1$; if $r \geq s_{\max}$, instead, there are no all-zero columns and $d\geq 2$. 

\end{IEEEproof}
In other words, Lemma \ref{lem:lemr2} states that $r(2)=s_{\max}$, i.e., that $n(2)=k+s_{\max}$. 

\begin{Lem}
    In a \ac{PRC-LDPC} code of block length $n(2)$, there are two families of codewords of weight $2$:
    \begin{itemize}
        \item one such that  the separation between the non-zero symbols is $\sum_{i=0}^{i^*}s_i$, called Family 1,
        \item the other such that  the separation between the non-zero symbols is $\sum_{i=i^*}^{w_h-2}s_i$, called Family 2,
    \end{itemize} where $i^*$ is the index for which $s_{i^*}=s_{\max}$. 
    \label{lem:famiglia}
\end{Lem}
\begin{IEEEproof}
Suppose again that the entries of $\6h$ separated by $s_{\max}$ are $h_i$ and $h_j$, $j>i$. Given that the \ac{RCC} is fulfilled, there can be only one pair of this kind. By definition, the closest non-zero symbol to the left of $h_j$ is $h_i$, and it holds that $j-i=s_{\max}$. When $n=n(2)$, implying $r=s_{\max}$, we have that $h_{0,i}=h_{1,i+1}=\ldots,h_{s_{\max}-1,i+s_{\max}-1}=h_i=1$. Since $j>i+s_{\max}-1$, we have that the only non-zero element in the $j$-th column is $h_{1,j}$. Therefore, the $0$-th and the $j$-th column of $\6H$ are equal, implying the existence of a codeword of weight $2$, where the two symbols one are separated by $j=s_0+\ldots+s_{\max}$ positions, proving the thesis. The second part of the lemma is proved with identical arguments, by considering the $i$-th and the last column of $\6H$.
\end{IEEEproof}

By performing polynomial division between $x^N+1$ and $h^*(x)$, it can be readily noticed that the two families described in Lemma \ref{lem:famiglia} appear consecutively in $\6p$.

The following straightforward  result holds for general values of $d$.

\begin{Lem}
    Consider any codeword $\6c$ of weight $w$ of a \ac{PRC-LDPC} code $\mathcal{C}$ with length $n$. Also consider the \ac{PRC-LDPC} code with length $n'=n-1$ obtained by puncturing $\mathcal{C}$, called $\mathcal{C}'$. Then, the vector obtained by removing the last entry of  $\6c$ is a codeword for $\mathcal{C}'$.
    \label{lem:puntu}
\end{Lem}
\begin{IEEEproof}
    It is a straightforward consequence of Remark \ref{rem:sliwi}.
\end{IEEEproof}

Lemma \ref{lem:puntu} implies that, when puncturing a symbol of any \ac{PRC-LDPC} code, either the minimum distance is preserved, or it decreases by at most one unit.

In the following we prove some other useful results on the weight distribution. According to the definition of family of codewords, we can state that 
\[
A(w)=\sum_{i} A_i(w),
\]
where $A_i(w)$ is the number of codewords of weight $w$ belonging to the $i$-th family. The following result holds.

\begin{The}
    If the \ac{PRC-LDPC} code of length $n$ is characterized by a certain $A_i(w)=A$, for any $i$, then the \ac{PRC-LDPC} code of length $n'=n+1$ has either $A_i(w)=A$, or $A_i(w)=\max\{A-1,0\}$, or $A_i(w)=A+1$.
\end{The}
\begin{IEEEproof}
As evident from \eqref{eq:trapezio} and the related discussion, a unitary increase in $n$ cannot cause the number of codewords in a given family to increase or decrease by more than one unit.
\end{IEEEproof}

\begin{The}
    If, for a certain value of $i$,  the \ac{PRC-LDPC} code of length $n$, called $\mathcal{C}$, exhibits $A_{i}(d)=1$, then the \ac{PRC-LDPC} code of length $n'=n+1$, called $\mathcal{C}'$, has $A(d+1)\geq 2$. If we call $\6c$ the codeword of $\mathcal{C}$ belonging to family $i$ and having weight $d$, then $\mathcal{C}'$ contains $[\6c,1]$ and $[1, \6c]$ as codewords. 
    \label{the:lemmaparoline}
\end{The}
\begin{IEEEproof}
   Under the assumption that we have only one codeword of weight $d$ left in the $i$-th family, called $\6c$, then in $\6p$ we have
   
\[\ldots,1,\tikz[baseline={(X.base)}]{
  \node[draw=black, rectangle, inner sep=1.5pt] (X) {0,\ldots,0};
},
1,\ldots
\]
where the symbols in the rectangle are $\6c$.  It is very easy to show that any other configuration breaks either the definition of minimum distance or the assumption that the codeword is unique for the $i$-th family. Given this, when considering a larger code length $n'=n+1$, then the larger window will cover the symbol one on the right, forming a codeword of weight $d+1$. Similarly, the symbol one on the left will be covered, forming a codeword of weight $d+1$, as well. Both these codewords belong to different families than $\6c$, since they include a larger number of symbols one. This proves the thesis.
\end{IEEEproof}

\begin{Cor}
    If, for a certain value of $i$,  the \ac{PRC-LDPC} code of block length $n$ shows $A_{i}(d)=1$ then the \ac{PRC-LDPC} code of length $n'=n+2$ has $A(d+2)\geq 1$. If we call $\6c$ the codeword of $\mathcal{C}$, then $\mathcal{C}'$ contains $[1,\6c,1]$ as a codeword.
\end{Cor}
\begin{IEEEproof}
    Same as Theorem \ref{the:lemmaparoline}, but the window of size $n+2$ covers both the symbols $1$ around $\6c$.
\end{IEEEproof}

All the above results can be used to find the low-weight codewords of \ac{PRC-LDPC} codes. In particular, our method relies on searching low-weight codewords in the portions of $\6p$ that we have shown to be sparse. Some more quantitative examples are provided next, where we choose  parity-check polynomials with a relatively small degree $k$. Then, we compute the weight distribution of the resulting family of \ac{PRC-LDPC} codes, validating the theoretical analysis. The reason for the choice of small values of $k$ is twofold: on the one hand, they lead to a more comprehensible treatment; on the other hand, we want to compare our results with the actual minimum distance profile, which would be uncomputable for large values of $k$.

\begin{Exa}

Let us start from the family of \ac{PRC-LDPC} codes obtainable from the parity-check polynomial
\[
h(x)=1+x+x^5+x^{11}+x^{13},
\]
having reciprocal
\[
h^*(x)=1+x^2+x^8+x^{12}+x^{13}.
\]
We have that $\6e=[0,1,5,11,13]$ is a Golomb ruler, and separations vector is $\6s=[1,4,6,2]$. We thus notice that the conditions in Remark \ref{rem:rul_ext_other} are satisfied.

By performing polynomial division between $x^N+1$ and $h^*(x)$ we obtain the following portion of $\6p$
\[
\ldots |1111|00000000|\mathbf{10100000100011}|00|1010101| \ldots,
\]
where we identify $\6h^*$, marked in bold, surrounded by two all-zero portions of size $2(1+4+6)-13-1=8$ and $2(6+2)-13-1=2$, and two comb-like zones, where symbols one are distant by $s_0=1$ positions on the left and $s_{w_h-2}=2$ positions on the right. This corroborates the analysis in Section \ref{sec:rules} and, in particular, in Fig. \ref{fig:zonapettini}. The shown sub-sequence already tells us that  $n(6)>24$, but it is possible to study the distance profile even more in detail. The analysis of the low-weight codewords by studying the portion of $\6p$ described by Lemma \ref{lem:famiglia} can be found in Appendix \ref{sec:APPA}. The reasoning in Appendix \ref{sec:APPA} already shows that the weight distribution of \ac{PRC-LDPC} codes, at least for relatively small values of $n$, can be characterized starting from a very thin portion of $\6p$. In this particular case, such a portion of $\6p$ is that defined by Lemma \ref{lem:famiglia}. Let us slightly widen this portion, i.e., let us consider
\begin{align*}
    \ldots 00001010101010000100011000000000010000000101001011\ldots
\end{align*}
It is interesting to notice that all the minimum weight codewords of all the \ac{PRC-LDPC} codes with $19 \leq n \leq37$, except for one of them when $n=21$, are contained into this portion of $\6p$, whose length is only $0.6\%$ of the total length of the pseudo-noise sequence, i.e., $2^k-1$.

\label{ex:exa2}

\end{Exa} 

\begin{figure}
    \centering
    \resizebox{!}{8cm}{
\begin{tikzpicture}

\begin{axis}[
  axis x line*=top,
  axis y line=none,
  xmin=0.2549, 
  xmax=0.72,
  xlabel={$\textcolor{white}{R'}R\textcolor{white}{R'}$},
  x dir=reverse,
 xticklabel style={
        /pgf/number format/fixed,
        /pgf/number format/precision=5
},
xtick style={
    /pgfplots/major tick length=0pt,
    /pgfplots/minor tick length=0pt
  }
]

\addplot[white]coordinates{(0.7,10)};
\end{axis}

  \begin{axis}[
ymin = 0,
ymax= 15,
grid = both,
xmin=18,
xmax=51,
legend style={at={(1.02,1)},anchor=north west},
mark size=2pt,
xlabel={$\textcolor{white}{R'}n\textcolor{white}{R'}$},
ylabel={$A(d)$},
xticklabel style={
        /pgf/number format/fixed,
        /pgf/number format/precision=5
},
ytick={1,2,3,4,5,6,7,8,9,10,11,12,13,14,15},
    yticklabels={$1$,$2$,$3$,$4$,$5$,$6$,$7$,$8$,$9$,$10$,$11$,$12$,$13$,$14$,$15$},
yticklabel style={
        font=\footnotesize
    } 
]

\addplot[black]coordinates{
(19,3)
(20,3)
(20,1)
(21,1)
(21,	8)
(22,8)
(22	,4)
(23,4)
(23	,2)
(24,2)
(24,	10)
(25,10)
(25	,6)
(26,6)
(26,	3)
(27,3)
(27	,2)
(28,2)
(28,	1)
(29,1)
(29	,4)
(30,4)
(30,	2)
(31,2)
(31	,7)
(32,7)
(32,	4)
(33,4)
(33,	1)
(34,1)
(34	,6)
(35,6)
(35,	2)
(36,2)
(36	,7)
(37,7)
(37,	2)
(38,2)
(38	,12)
(39,12)
(39,	6)
(40,6)
(40	,2)
(41,2)
(41,	1)
(42,1)
(42,	3)
(43,3)
(43,	1)
(44,1)
(44,	4)
(45,4)
(45,	2)
(46,2)
(46,	1)
(47,1)
(47,	3)
(48,3)
(48,	1)
(49,1)
(49,	2)
(50,2)
(50,	6)
};\addlegendentry{Actual};

\addplot[black, mark= x, only marks]coordinates{
(19,3)
(20,1)
(21,	7)
(22	,4)
(23	,2)
(24,	10)
(25	,6)
(26,	3)
(27	,2)
(28,	1)
(29	,4)
(30	,2)
(31	,7)
(32,	4)
(33,	1)
(34	,6)
(35,	2)
(36	,7)
(37,	2)
(38	,10)
(39,	5)
(40	,2)
(41,	1)
(42,	3)
(43,	1)
(44,	4)
(45,	2)
(46,	1)
(47,	3)
(48,	1)
(49,	2)
(50,	3)
};\addlegendentry{Estimate};

\addplot[red, mark=x, only marks]coordinates{
(21,	4.52797317504883)
(22,	2.41350650787354)
(23,	1.23998963832855)
(24,	4.56665968894959)
(25,	2.55175864696503)
(26,	1.38798376917839)
(27,	0.731382742524147)
(28,	0.370404571294785)
(29,	1.42207200825214)
(30,	0.784439669921994)
(31,	2.37388394074515)
(32,	1.38087298348546)
(33,	0.787046688608825)
(34,	2.17565744929016)
(35,	1.29410742013715)
(36,	3.17766599706374)
(37,	1.95332254108507)
(38,	4.38946736787329)
(39,	2.77188493435824)
(40,	1.72749078212655)
(41,	1.06203153249317)
(42,	2.39562236786719)
(43,	1.51054419860213)
(44,	3.19057845942007)
(45,2.05576672142615)
(46,	1.30935011461060)
(47,	2.70075400733025)
(48,	1.75434030726981)
(49,	3.44684056185210)
(50	,6.33303980592488)
};\addlegendentry{Average [14]};


\end{axis}

\end{tikzpicture}}
    \caption{Comparison of $A(d)$ for the \ac{PRC-LDPC} codes in Example \ref{ex:exa2}, with $k=13$.}
    \label{fig:Ad}
\end{figure}
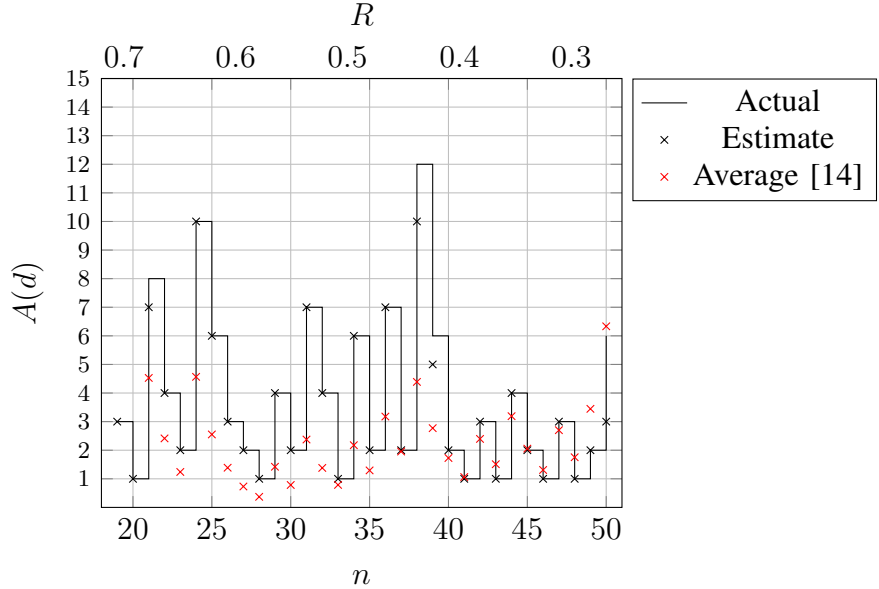
\begin{figure}
    \centering
    \resizebox{!}{8cm}{
\begin{tikzpicture}
  \begin{axis}[
ytick={18,19,20,21,22,23,24,25,26,27,28,29,30,31,32,33,34,35},        
yticklabels={,$19$,$20$,$21$,$22$,$23$,$24$,$25$,$26$,$27$,$28$,$29$,$30$,$31$,$32$,$33$,$34$,$35$},        
ymin = 18,
ymax= 35,
grid = both,
xmin=1,
xmax=8,
legend style={at={(1.02,1)},anchor=north west},
mark size=2pt,
xlabel={$\textcolor{white}{R'}d\textcolor{white}{R'}$},
ylabel={$n(d)$},
xticklabel style={
        /pgf/number format/fixed,
        /pgf/number format/precision=5
},
yticklabel style={
        font=\footnotesize
    }]

\addplot[black, line width=0.7pt]coordinates{
(3, 19)
(4,19)
(4,21)
(5,21)
(5,24)
(6,24)
(6,29)
(7,29)
(7,31)
};\addlegendentry{$k=13$, actual};

\addplot[black, mark = x, only marks]coordinates{
(3, 19)
(4,21)
(5,24)
(6,29)
(7,31)
};\addlegendentry{$k=13$, estimate};

\addplot[red, line width=0.7pt]coordinates{
(2, 21)
(3,21)
(3,23)
(4,23)
(4,28)
(5,28)
(5,30)
(6,30)
(6,31)
(7,31)
(7,34)
};\addlegendentry{$k=15$, actual};


 \addplot[red, mark = o, only marks]coordinates{
(2, 21)
(3,23)
(4,28)
(5,30)
(6,31)
(7,33)
};\addlegendentry{$k=15$, estimate};

\end{axis}

\end{tikzpicture}}
    \caption{Minimum distance profile of the \ac{PRC-LDPC} codes in Examples \ref{ex:exa2} and \ref{ex:exa3}.}
    \label{fig:samples_vs_m}
\end{figure}
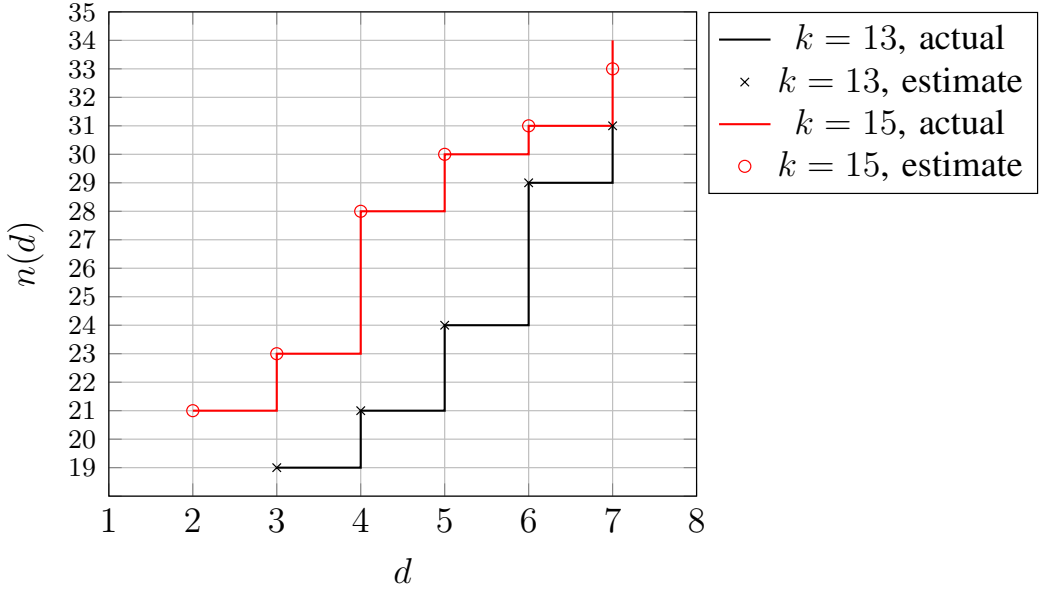

In Fig. \ref{fig:Ad} we compare the actual $A(d)$ of the codes in Example \ref{ex:exa2}, for different values of $n$, to that obtained using our method (looking around the small portion of $\6p$ specified by Lemma \ref{lem:famiglia}) and also to the average value, computed as in \cite{Shirvanimoghaddam}.  We notice that in the considered range $0.26 \leq R \leq 0.68$, our estimate perfectly matches the actual value of $A(d)$, except for a handful of cases. We remark that the values of $d$, which depend on $n$, are immediately deducible from Fig. \ref{fig:samples_vs_m}, introduced in Example \ref{ex:exa3}. Namely, $d=3$ when $n=19$, and then each increase of $A(d)$ corresponds to a unitary increase of $d$. The average estimator (working for $d\geq 3$) provides good results for relatively large values of $n$, and gives less precise estimates when $n$ is small.

\begin{Exa}

We also consider the family of \ac{PRC-LDPC} codes obtainable from 
\[
h(x)=1+x^2+x^8+x^{12}+x^{15},
\]
which implies $\mathbf{e}=[0,2,8,12,15]$, and then $\mathbf{s}=[2,6,4,3]$.

For the sake of brevity, in this example we do not list all the codewords as in Example \ref{ex:exa2} (and, in particular, in Appendix \ref{sec:APPA}), since the rationale is exactly the same. We only mention that the portion of $\6p$ around the families introduced in Lemma \ref{lem:famiglia}, which is
\begin{align*}
\ldots 011100001010000000100000000000010110\ldots,
\end{align*}
contains all the minimum distance codewords of the \ac{PRC-LDPC} codes with $$n=22,26,27,28,29.$$ More in general, somehow surprisingly, this portion of $\6p$ consists of $0.1\%$ of the total length of $\6p$ and contains $59\%$ of the minimum distance codewords of all the \ac{PRC-LDPC} codes derived from the above $h(x)$, with $21\leq n \leq 33$.















\label{ex:exa3}
\end{Exa}

We show  in Fig. \ref{fig:samples_vs_m} the initial values of $n(d)$ for the codes in Examples \ref{ex:exa2} and \ref{ex:exa3}. It is also  remarkable that, if the only goal of the analysis is to estimate the minimum distance (and not enumerating the number of minimum distance codewords), our method basically provides exact matching in all the considered cases, except for a single discrepancy when $k=15$ and $n=33$, where the estimated distance is $7$, rather than $6$. We remark that, for the same value of $n$, the code with $k=13$ has lower code rate than the code with $k=15$.

\subsection{Code design}

In this section we describe the method we employ to find parity-check polynomials fulfilling the properties described in Section \ref{sec:sequense} and summarized in Fig. \ref{fig:Venn}. 

Let us suppose that an $(n,k)$ code must be obtained and the target row-weight of the parity-check matrix is $w_h$. The design we adopt is as follows:
\begin{enumerate}
    \item Pick a Golomb ruler, say $\mathcal{G}$, of order $L\gg w_h$, containing $k$ among its marks.\footnote{$\mathcal{G}$ can be either a known Golomb ruler or one designed ad hoc. Golomb rulers with up to $65000$ marks have been exhaustively designed and tabulated (see \cite{Golombrulers}). It is extremely unlikely that there exist practical values of the code dimension $k$ that are not a mark in any of them. In any case, it is always possible to start from $k'>k$ and then adjust $k'$ and the code rate by combining puncturing and shortening operations.} 
    \item Remove all the marks larger than $k$, so obtaining a new Golomb ruler, say $\mathcal{G}'$, of order $L'$. If $L'<w_h$, go back to Step 1). Set $j=0$.
    \item Pick the $j$-th subset of size $w_f<w_h$ of the marks of $\mathcal{G}'$, say $S_f^{(j)}$.  If the polynomial associated to these marks breaks the conditions in Remark \ref{rem:rul_ext_other},  set $j=j+1$ and restart this step.  Set $i=0$.
    \item Let us call $S_r^{(j)}$ the complementary set of  $S_f^{(j)}$. Pick the $i$-th  subset of $S_r^{(j)}$ of size $w_h-w_f$, called $S_r^{(j,i)}$.
    \item Test whether the polynomial $$p^{(j,i)}(x)=\sum_{l \in \{S_f^{(j)} \cup S_r^{(j,i)}\}} x^l$$ is primitive and store $p^{(j,i)}(x)$ if it is. If $i=\binom{L'-w_f}{w_h-w_f}-1$, set $j=j+1$ and go back to Step 3), else set $i=i+1$ and go back to Step 4).
\end{enumerate}

This procedure generates all the primitive polynomials, if any, associated to a certain Golomb ruler $\mathcal{G'}$. It requires testing at most $\binom{L'}{w_f}\binom{L'-w_f}{w_h-w_f}$ polynomials for primitivity. 
It is evident that when the objective is to generate a single family of \ac{PRC-LDPC} codes, the algorithm can be halted as soon as the first parity-check polynomial $p^{(j,i)}(x)$ is found. 

$\mathcal{G}$ (and thus $L'$) and $w_f$ should be regarded as degrees of freedom in code design, as their selection impacts the trade-off between search complexity and the probability of finding a primitive polynomial within the search space.

\section{Encoding and decoding complexity}
\label{sec:complex}

\ac{PRC-LDPC} codes can be encoded by using an encoder circuit of the type in Fig. \ref{fig:encoder}. To be noted that such a circuit is basically a \ac{LFSR} with characteristic polynomial $h^*(x)$, according to Property \ref{pro:prognoise}. The reason for including the switches for the coefficients of $h^*(x)$ is that this makes the encoder circuit suitable for \ac{PRC-LDPC} codes of any block length smaller than or equal to $2^k-1$, described by any parity-check polynomial. In other words, the  encoder circuit in Fig. \ref{fig:encoder} can be employed to encode any \ac{PRC-LDPC} code described by a parity-check polynomial of degree smaller than or equal to $k$. The polynomial $i(x)$ represents the binary sequence generated by the information source. The polynomial $c(x)$, whose degree is at most $r-1$, represents the sequence of parity-check symbols. As shown in Fig. \ref{fig:encoder}, the rightmost switch stays up for the first $k$ time instants and down for the remaining $r$ time instants, in order to form a codeword of length $n$. The other switches are closed only if the corresponding coefficients of $h^*_i$ is $1$. Therefore, the computational complexity of encoding a codeword of a \ac{PRC-LDPC} code is the same as that of generating a $n$-bit sequence using an \ac{LFSR}, i.e., $O(n)$.

Such an encoding complexity is clearly smaller than the usual $O(n^2)$ complexity of general systematic encoding \cite{Nguyen2019}, based on multiplication of the information vector by the code generator matrix (which, in turn, can be derived from $\6H$ through Gaussian elimination, costing $O(n^3)$).  
In the \ac{RU} method \cite{Richardson2001}, according to which the parity-check matrix is brought into approximate lower triangular form, encoding is performed without using the generator matrix. The complexity of the encoding algorithm is  $O(n+g^2)$, where $g$ is the number number of rows of $\6H$ that cannot be brought into triangular form by using only row and
column permutations and is usually in the order of $\sqrt{n}$. It is well-known that also \ac{QC-LDPC} codes can be encoded by using shift registers, as reported, for example, in \cite{Costello}. Many implementations of encoder design for \ac{QC-LDPC} codes have been proposed, after the seminal work \cite{Zongwang2006}. However, these encoders usually require more than one\footnote{The exact number depends on the chosen implementation.} shift register \cite[Fig. 3]{Zongwang2006}, due to the fact that, differently from our codes, \ac{QC-LDPC} codes are not characterized by a single  unidimensional structure, but rather by a bidimensional one (in general, they are represented by a base matrix, rather than a polynomial). 


\begin{figure}[thb]
\centering
\begin{tikzpicture}[>=stealth, node distance=1.5cm, every node/.style={draw}]
  
  \node (rect1) [rectangle, minimum size=1cm] {$(1)$};
  \node (rect2) [rectangle, right=of rect1, minimum size=1cm] {$(2)$};
  \node (rect3) [rectangle, draw=none, right=of rect2,minimum size=1cm] {$\cdots$};
  \node (rect4) [rectangle, right=of rect3,minimum size=1cm] {$(k)$};

   \node (circle1) [circle, fill=white, inner sep=1pt, minimum size=0.2cm, right=of rect4] {};
   \draw (circle1) -- (10.4,-0.2);
   
 \node (circle2) [circle, fill=white, inner sep=1pt, minimum size=0.2cm, above right=0.3cm and 0.5cm of circle1] {};
 \node (circle3) [circle,node distance=1cm, fill=white, inner sep=3pt, minimum size=0.5cm, right= of circle2] {S};
 \node (circle4) [circle, fill=white, inner sep=1pt, minimum size=0.2cm, above right=-0.7
 cm and 0.5cm of circle1] {};

  \draw [<-] (rect1.east) -- (rect2.west);
  \draw [<-] (rect2.east) -- (rect3.west);
  \draw [<-] (rect3.east) -- (rect4.west);
  
   \draw [->] (circle1.west) -- (rect4.east);
      \draw [<-] (circle2.east) -- (circle3.west) node [font=\footnotesize,above, draw= none, midway,pos=.7] {$i(x)$} node [font=\tiny,below =0.8cm of circle3, draw= none, midway, pos=1.3] {$k,\ldots,k+r-1$};

   \coordinate (midpoint1) at ($(circle1.west)!0.5!(rect4.east)$);
   
\draw [->] (midpoint1) -- ++(0,0.8cm) -- ++(-11cm,0) node [font=\footnotesize, above, draw= none, midway, pos=0.8] {$t(x)=x^ri(x)+c(x)$} node [font=\tiny,right, draw= none, midway, pos=-0.01] {$0,1,\ldots,k-1$};

 \coordinate (midpoint2) at ($(rect1.west)$);
   \draw [->] (midpoint2) -- ++(-0.5cm,0) -- ++(0,-1cm);

\coordinate (midpoint3) at ($(rect1.east)!0.5!(rect2.west)$);
   \draw [->] (midpoint3)  -- ++(0,-1cm);

\coordinate (midpoint4) at ($(rect3.east)!0.5!(rect4.west)$);
   \draw [->] (midpoint4)  -- ++(0,-1cm);

   \node (circle5) [circle, fill=white, inner sep=1pt, minimum size=0.2cm,node distance=1cm, below= of midpoint4] {};
   \node (circle6) [circle, fill=white, inner sep=1pt, minimum size=0.2cm,node distance=1cm, below= of midpoint3] {};
   \node (circle7) [circle, fill=white, inner sep=1pt, minimum size=0.2cm,node distance=2.05cm, left= of circle6] {};

   \node (circle8) [circle, fill=white, inner sep=1pt, minimum size=0.2cm,node distance=0.5cm, below= of circle5] {};
   \node (circle9) [circle, fill=white, inner sep=1pt, minimum size=0.2cm,node distance=0.5cm, below= of circle6] {};
     \node (circle10) [circle, fill=white, inner sep=1pt, minimum size=0.2cm,node distance=0.5cm, below= of circle7] {};
 \node (circlefake) [circle, fill=white, inner sep=1pt, draw=none, minimum size=0.2cm,node distance=2.05cm, right= of circle9] {$\cdots$};

\draw (circle10) -- (-0.5,-1.3);

\draw (circle9) -- (1.7,-1.3);

\draw (circle8) -- (6.8,-1.3);
 
       \node (circle11) [circle, fill=white, inner sep=3pt, minimum size=0.5cm,node distance=1.5cm, below right= 1.9 cm and 0.5cm of rect4] {$+$};

\draw [->] (circle10.south) -- ++(0,-1.3cm) -- ++(9.82,0) --(circle11.south);
\draw [->] (circle9.south) -- ++(0,-0.745) -- (circle11.west);
\draw [->] (circle8.south) -- ++(0,-0.25cm) -- ++(2.52cm,0) -- (circle11.north);

\draw [->] (circle11) -- ++(1.51cm,0cm) -- (circle4.south) node [font=\footnotesize, above, draw= none, left, pos=0.2] {$c(x)$};

\draw (-1,-1.4) ellipse (0.8cm and 0.7cm) node  [draw=none,  shift={(-2,0)}] {$h^*_k=h_0=1$};
\draw (1.25,-1.4) ellipse (0.8cm and 0.7cm) node  [draw=none, shift={(1.8,0)}] {$h^*_{k-1}=h_1$};
\draw (6.25,-1.4) ellipse (0.8cm and 0.7cm) node [draw=none,  shift={(1.8,0)}] {$h^*_1=h_{k-1}$};

\end{tikzpicture}
\caption{Encoder circuit for \ac{PRC-LDPC} codes.}
\label{fig:encoder}
\end{figure}

We remark that the encoder circuit we employ is inspired by that in \cite{Felstrom1999}. However, differently from \cite{Felstrom1999}, where time varying convolutional codes are treated, we employ the encoder in Fig. \ref{fig:encoder} to encode block codes. This difference implies that the switch on the right side of Fig. \ref{fig:encoder} needs to commute with a different rate than in the convolutional case. However, it is interesting to notice that \ac{PRC-LDPC} codes share some structural properties with \ac{LDPC} convolutional codes. For example, in both cases the parity-check matrix has a diagonal-like structure. As a spark for future works, we mention that a convolutional version of \ac{PRC-LDPC} codes can be obtained by applying a different type of puncturing, called constant-length puncturing \cite{Cancellieri2015}. If this operation is applied, the resulting parity-check matrix has the same number of columns as the initial simplex code, but a smaller number of rows. Then, an infinite frame has to be constructed. However, the analysis of convolutional \ac{PRC-LDPC} codes requires a self-standing approach and, for this reason, we do not delve into them in this paper.

Regarding decoding complexity, the sparsity of the parity-check matrix of \ac{PRC-LDPC} codes enables the employment of low-complexity \ac{BP}-based decoding algorithms, such as the \ac{SPA}, \ac{MS}, \ac{NMS}, and many others. Let us consider the implementation of the \ac{LLR-SPA} decoder proposed in \cite{Hu2001} (which is also the one we consider for the Monte Carlo simulations reported in Section \ref{sec:montecarlo}) and operations between $8$-bit values, we have that the number of binary operations required to decode a codeword is
\begin{equation}
\Lambda=nI_{\mathrm{avg}}f(\langle w_c \rangle,R), 
\label{eq:complexitySPA}
\end{equation}
where
\[
f(\langle w_c \rangle,R)=8(8\langle w_c \rangle+12R-11)+\langle w_c \rangle,
\]
and $I_{\mathrm{avg}}$ is the average number of decoding iterations. By substituting \eqref{eq:wcmedio} into \eqref{eq:complexitySPA}, we get that $\Lambda$ is $O(nw_h)$.

\section{Performance assessment}
\label{sec:montecarlo}
We assess the performance of some short codes, in terms of \ac{CER}, through Monte Carlo simulations of \ac{BPSK} modulated transmissions over the \ac{AWGN} channel. We compare the performance of a \ac{PRC-LDPC} code characterized by $\6e=(0,2,21,29,60,72,75)$ with state-of-the-art short \ac{LDPC} codes with $R=\frac{1}{2}$ and $n=128$ \cite{Livashort}, namely
\begin{itemize}
    \item an AR3A LDPC code \cite{DIVSALARAR3A},
    \item an ARJA LDPC code \cite{ARJA},
    \item the CCSDS telecommand LDPC code based on protographs \cite{Divsalar},
    \item a standard  $(3,6)$-regular LDPC code.
\end{itemize}
The \ac{PRC-LDPC} code, for which $\6s=(2,19,8,31,12,3)$, indeed verifies Remark \ref{rem:rul_ext_other}. Moreover, $s_{\max}$ is such that Corollary \ref{cor:distanza_finestra} does not hold. For $R=\frac{1}{2}$, i.e., $n=150$, the estimated minimum distance of this code (obtained by using the tool in \cite{MacKayDist}) is $11$.
In order to have the exact same length of the other codes, we have applied shortening operations on the \ac{PRC-LDPC} code, by considering $z=11$ and by puncturing $11$ more symbols, in such a way that $k'=k-z=64$, and $n'=n-z-11=128$. The estimated minimum distance of the shortened code is $10$. We note that its performance is comparable to that of the other considered codes. We have verified that the loss of the shortened code with respect to the original code with block length $n=150$ is always smaller than $0.2$ dBs. Clearly, there is a trade-off between  the encoding/decoding complexity and the error rate performance, and the slight performance loss we incur in with respect to the AR3A and ARJA codes is justified by the employment of an extremely simple encoding circuit, as discussed in Section \ref{sec:complex}.

 
\begin{figure}[tb]
    \centering
    \includegraphics[keepaspectratio, width=10cm]{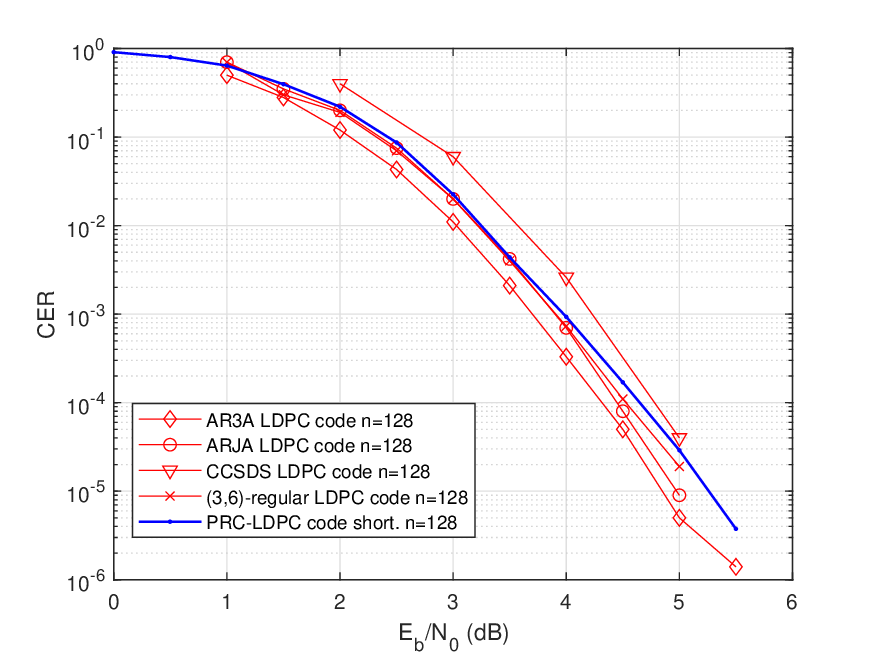}
    \caption{Codeword Error Rate vs $E_b/N_0$, for various codes in comparison with \ac{PRC-LDPC} codes under LLR-SPA decoding.}
    \label{fig:HCER}
    \vspace{-1.7em}
\end{figure}

\section{Conclusion}
\label{sec:conclusions}
A new family of LDPC codes, obtained by puncturing cyclic simplex codes and called \ac{PRC-LDPC} codes, has been proposed. These codes exhibit several benefits, including simple encoders, the possibility to be decoded with low-complexity algorithms and inherent rate compatibility. 
Their encoder is a one-of-a-kind device that, for a chosen parity-check polynomial of degree $k$, can generate codes of variable length up to $2^k-1$, and the connections representing the reciprocal of the parity-check polynomial can be configured by simply toggling certain switches. 
As another advantage, the minimum distance properties of \ac{PRC-LDPC} codes can be analyzed and estimated using theoretical methods, especially for moderate to high values of the code rate.

\appendices

\section{Proof of Lemma \ref{lem:whprimo} \label{app:appendiceprova}}

 When $R=\frac{1}{2}$, $\6H$ contains $s_0+s_{w_h-2}$  columns of weight $w_h-1$ in its central part and $s_0+s_{w_h-2}$ columns of weight $1$ at the right and left sides, each representing a circularly shifted version of $\6h^*$. The central columns circularly shift down by one position (from left to right) and, therefore, the support of any of them can be easily obtained from the support of the first one by computing a sum modulo $k$ (that for $R=1/2$ is equal to $r$). Instead, the $i$-th column, for $i\in[1,s_0]$, has support exactly equal to $i$ and, for $i\in[n-s_{w_h-2}+1,n]$, the support of the $i$-th column is exactly $r-n+i$. This implies that, if the support of any column of weight $w_h-1$ takes entries in $[1,s_0]\cup[r-s_{w_h-2}+1,r]$, a codeword is generated, since there are $w_h$ columns summing to zero modulo $2$ ($w_h-1$ of weight $1$ and $1$ of weight $w_h-1$). We call the latter statement \emph{Condition 1}. A sufficient condition for the existence of these columns is that, for either $i=0$ and/or $i=w_h-2$
 $$s_i>\sum_{j\in[1,w_h-3]}s_j,$$
ensuring that the $w_h-1$ ones forming the $s_j$'s with $j\in[1,w_h-3]$ are grouped in less than $s_0+s_{w_h-2}$ rows (modulo $r$). Therefore, if the first symbol one forming $s_0$ (or, equivalently, $s_{w_h-2}$) has row index $\in[s_0-\sum_{j\in[1,w_h-3]}s_j,s_1]$, then by hypothesis the second symbol one forming $s_0$ (or, equivalently, $s_{w_h-2}$) has row index in $[r-s_{w_h-2}+1,r]$, all the symbols in $[s_0+1,r-s_{w_h-2}]$ are zeros and Condition 1 holds.
Clearly, due to the cyclic shifts, this condition occurs $s_0-\sum_{j\in[1,w_h-3]}s_j$ times. Obviously, if $s_{w_h-2}>\sum_{j\in[1,w_h-3]}s_j$, then $s_{w_h-2}$ substitutes $s_0$ in the previous equation.

\section{Low-weight codewords of the \ac{PRC-LDPC} codes in Example \ref{ex:exa2}} \label{sec:APPA}

Given the parity-check polynomial
\[
h(x)=1+x+x^5+x^{11}+x^{13},
\]
we have that $\6e=[0,1,5,11,13]$ is a Golomb ruler, and the following vector containing separations is  
$\6s=[1,4,6,2]$. We study next the portion of $\6p$ described by Lemma \ref{lem:famiglia}, showing how low-weight codewords arise for increasing values of $n$.

For this family of \ac{PRC-LDPC} codes, we have $r(2)=6$ and therefore $n(2)=19$.  By exhaustive search we find $A(2)=3$, and these three codewords indeed belong to Families 1 ($2$ out of $3$) and 2 (the remaining $1$), as described in Lemma  \ref{lem:famiglia}. They are marked in bold in the following portion of $\6p$

\begin{align*}
\ldots10001|\textbf{1000000000010000000}|101001\ldots\\
\ldots100011|\textbf{0000000000100000001}|01001\ldots\\
\ldots1000110|\textbf{0000000001000000010}|1001\ldots\\
\end{align*}
By increasing $n$ to $20$, we get $A(2)=1$, still belonging to Family 2 
\begin{align*}
\ldots100011|\textbf{00000000001000000010}|1001\ldots
\end{align*}
According to Corollary \ref{cor:corsidon}, we have $n=21=n(3)$ and indeed $d=3$ and $A(d)=8$, where the following codewords are derived from Families 1 and 2 as described in Lemma \ref{lem:puntu} (applied for increasing, rather than decreasing, values of $n$), 
\begin{align*}
\ldots10001|\textbf{100000000001000000010}|1001\ldots\\
\ldots100011|\textbf{000000000010000000101}|001\ldots
\end{align*}
and the others belong to new families. We remark that all the $8$ codewords of weight $3$, except for one of them, can be found in the above portion of $\6p$.  Finally, when $n=22$, the two aforementioned codewords merge in a single codeword of weight $4$ 
\begin{align*}
\ldots10001|\textbf{1000000000010000000101}|001\ldots
\end{align*}
which, however it is not a minimum distance codeword, since $A(3)=4$ and, again, all the minimum codewords come from the shown portion of $\6p$. The \ac{PRC-LDPC} code instead assumes $d=4$ when $n=24$, for which $A(4)=10$. For the sake of brevity, we only show the four codewords of weight $4$ deriving from the straightforward application of the results in Section \ref{sec:rules} to the codewords listed above but, once more, all the codewords of weight $4$ can be found in the portion of $\6p$ shown below
\begin{align*}
\ldots1|\textbf{000110000000000100000001}|01001\ldots\\
\ldots|\textbf{100011000000000010000000}|101001\ldots\\
\ldots100011|\textbf{000000000010000000101001}|\ldots\\
\ldots10001|\textbf{100000000001000000010100}|1\ldots
\end{align*}


\bibliographystyle{IEEEtran}
\bibliography{Archive}

\begin{thebibliography}{10}
\providecommand{\url}[1]{#1}
\csname url@samestyle\endcsname
\providecommand{\newblock}{\relax}
\providecommand{\bibinfo}[2]{#2}
\providecommand{\BIBentrySTDinterwordspacing}{\spaceskip=0pt\relax}
\providecommand{\BIBentryALTinterwordstretchfactor}{4}
\providecommand{\BIBentryALTinterwordspacing}{\spaceskip=\fontdimen2\font plus
\BIBentryALTinterwordstretchfactor\fontdimen3\font minus
  \fontdimen4\font\relax}
\providecommand{\BIBforeignlanguage}[2]{{%
\expandafter\ifx\csname l@#1\endcsname\relax
\typeout{** WARNING: IEEEtran.bst: No hyphenation pattern has been}%
\typeout{** loaded for the language `#1'. Using the pattern for}%
\typeout{** the default language instead.}%
\else
\language=\csname l@#1\endcsname
\fi
#2}}
\providecommand{\BIBdecl}{\relax}
\BIBdecl

\bibitem{Battaglioni2022Fitce}
M.~Battaglioni and G.~Cancellieri, ``Punctured binary simplex codes as {LDPC}
  codes,'' in \emph{2022 61st FITCE International Congress Future
  Telecommunications: Infrastructure and Sustainability (FITCE)}, 2022, pp.
  1--6.

\bibitem{Battaglioni2023ICC}
M.~Battaglioni, M.~Baldi, F.~Chiaraluce, and G.~Cancellieri, ``{Rate-adaptive
  LDPC codes obtained from simplex codes},'' in \emph{{Proc. of 2023 IEEE
  International Conference on Communications (ICC)}}, 2023, pp. 1--6.

\bibitem{Ha2004}
J.~Ha, J.~Kim, and S.~McLaughlin, ``Rate-compatible puncturing of low-density
  parity-check codes,'' \emph{IEEE Transactions on Information Theory},
  vol.~50, no.~11, pp. 2824--2836, 2004.

\bibitem{DAVIDA1972}
\BIBentryALTinterwordspacing
G.~I. Davida and S.~M. Reddy, ``Forward-error correction with decision
  feedback,'' \emph{Information and Control}, vol.~21, no.~2, pp. 117--133,
  1972. [Online]. Available:
  \url{https://www.sciencedirect.com/science/article/pii/S0019995872900575}
\BIBentrySTDinterwordspacing

\bibitem{3GPP}
{3GPP}, ``{Flexibility evaluation of channel coding schemes for NR-Discussion
  and Decision},'' {3GPP TSG TSG RAN WG1 Meeting 86}, October 2016.

\bibitem{Hui2018}
D.~Hui, S.~Sandberg, Y.~Blankenship, M.~Andersson, and L.~Grosjean, ``{Channel
  coding in 5G New Radio: a tutorial overview and performance comparison with
  4G LTE},'' \emph{IEEE Vehicular Technology Magazine}, vol.~13, no.~4, pp.
  60--69, 2018.

\bibitem{RyanBook}
W.~E. Ryan and S.~Lin, \emph{Channel codes - Classical and modern}.\hskip 1em
  plus 0.5em minus 0.4em\relax New York: Cambridge University Press, 2009.

\bibitem{ElKhamy2009}
M.~El-Khamy, J.~Hou, and N.~Bhushan, ``{Design of rate-compatible structured
  LDPC codes for hybrid ARQ applications},'' \emph{IEEE Journal on Selected
  Areas in Communications}, vol.~27, no.~6, pp. 965--973, 2009.

\bibitem{Ivanov2017}
F.~I. Ivanov and P.~S. Rybin, ``On the nested family of {LDPC} codes based on
  {G}olomb rulers,'' in \emph{2017 IVth International Conference on Engineering
  and Telecommunication (EnT)}, 2017, pp. 67--71.

\bibitem{Xiao2021}
X.~Xiao, B.~Vasic, S.~Lin, J.~Li, and K.~Abdel-Ghaffar, ``Quasi-cyclic {LDPC}
  codes with parity-check matrices of column weight two or more for correcting
  phased bursts of erasures,'' \emph{IEEE Trans. on Commun.}, vol.~69, no.~5,
  pp. 2812--2823, 2021.

\bibitem{Kim2022}
I.~Kim and H.-Y. Song, ``A construction for girth-8 {QC}-{LDPC} codes using
  {G}olomb rulers,'' \emph{Electron. Lett.}, vol.~58, no.~15, pp. 582--584,
  2022.

\bibitem{Chen2012}
C.~Chen, B.~Bai, Z.~Li, X.~Yang, and L.~Li, ``Nonbinary cyclic {LDPC} codes
  derived from idempotents and modular {G}olomb rulers,'' \emph{IEEE Trans. on
  Commun.}, vol.~60, no.~3, pp. 661--668, 2012.

\bibitem{Zhao2016}
S.~Zhao and X.~Ma, ``Construction of high-performance array-based non-binary
  {LDPC} codes with moderate rates,'' \emph{IEEE Communications Letters},
  vol.~20, no.~1, pp. 13--16, 2016.

\bibitem{Shirvanimoghaddam}
M.~Shirvanimoghaddam, ``Primitive rateless codes,'' \emph{IEEE Trans. on
  Commun.}, vol.~69, no.~10, pp. 6395--6408, 2021.

\bibitem{Tanner1981}
M.~R. Tanner, ``A recursive approach to low complexity codes,'' \emph{IEEE
  Trans. Inf. Theory}, vol.~27, no.~5, pp. 533--547, 1981.

\bibitem{BaldiSimplex}
M.~Baldi, M.~Bianchi, F.~Chiaraluce, and T.~Klove, ``A class of punctured
  simplex codes which are proper for error detection,'' \emph{IEEE Trans. on
  Inf. Theory}, vol.~58, no.~6, pp. 3861--3880, 2012.

\bibitem{Kschischang}
F.~Kschischang, B.~Frey, and H.-A. Loeliger, ``Factor graphs and the
  sum-product algorithm,'' \emph{IEEE Trans. on Inf. Theory}, vol.~47, no.~2,
  pp. 498--519, 2001.

\bibitem{Golomb}
S.~W. Golomb, \emph{Digital communications with space applications}.\hskip 1em
  plus 0.5em minus 0.4em\relax Prentice-Hall, Inc., 1964.

\bibitem{Fredricsson1975}
S.~Fredricsson, ``Pseudo-randomness properties of binary shift register
  sequences (corresp.),'' \emph{IEEE Transactions on Information Theory},
  vol.~21, no.~1, pp. 115--120, 1975.

\bibitem{Golombrulers}
A.~Dimitromanolakis, ``{Golomb rulers and Sidon sets},''
  \url{http://www.cs.toronto.edu/~apostol/golomb/results}, accessed:
  05/19/2023.

\bibitem{Nguyen2019}
\BIBentryALTinterwordspacing
T.~T.~B. Nguyen, T.~Nguyen~Tan, and H.~Lee, ``{Efficient QC-LDPC encoder for 5G
  New Radio},'' \emph{Electronics}, vol.~8, no.~6, 2019. [Online]. Available:
  \url{https://www.mdpi.com/2079-9292/8/6/668}
\BIBentrySTDinterwordspacing

\bibitem{Richardson2001}
T.~Richardson and R.~Urbanke, ``{Efficient encoding of low-density parity-check
  codes},'' \emph{IEEE Transactions on Information Theory}, vol.~47, no.~2, pp.
  638--656, 2001.

\bibitem{Costello}
S.~Lin and D.~J. Costello, \emph{Error control coding (2nd Edition)}.\hskip 1em
  plus 0.5em minus 0.4em\relax Prentice-Hall, Inc., 2004.

\bibitem{Zongwang2006}
Z.~Li, L.~Chen, L.~Zeng, S.~Lin, and W.~Fong, ``{Efficient encoding of
  quasi-cyclic low-density parity-check codes},'' \emph{IEEE Transactions on
  Communications}, vol.~54, no.~1, pp. 71--81, 2006.

\bibitem{Felstrom1999}
A.~Jim\'{e}nez~Felstr\"{o}m and K.~S. Zigangirov, ``Time-varying periodic
  convolutional codes with low-density parity-check matrix,'' \emph{IEEE Trans.
  on Inf. Theory}, vol.~45, no.~6, pp. 2181--2191, Sep. 1999.

\bibitem{Cancellieri2015}
G.~Cancellieri, \emph{Polynomial theory of error correcting codes}.\hskip 1em
  plus 0.5em minus 0.4em\relax Springer, 2015.

\bibitem{Hu2001}
X.-Y. Hu, E.~Eleftheriou, D.-M. Arnold, and A.~Dholakia, ``{Efficient
  implementations of the sum-product algorithm for decoding LDPC codes},'' in
  \emph{GLOBECOM'01. IEEE Global Telecommunications Conference (Cat.
  No.01CH37270)}, vol.~2, 2001, pp. 1036--1036E vol.2.

\bibitem{Livashort}
\BIBentryALTinterwordspacing
G.~Liva, L.~Gaudio, T.~Ninacs, and T.~Jerkovits, ``Code design for short
  blocks: A survey,'' 2016. [Online]. Available:
  \url{https://arxiv.org/abs/1610.00873}
\BIBentrySTDinterwordspacing

\bibitem{DIVSALARAR3A}
D.~Divsalar, S.~Dolinar, and J.~Thorpe,
  ``Accumulate-repeat-accumulate-accumulate-codes,'' in \emph{IEEE 60th
  Vehicular Technology Conference, 2004. VTC2004-Fall. 2004}, vol.~3, 2004, pp.
  2292--2296.

\bibitem{ARJA}
D.~Divsalar, S.~Dolinar, C.~R. Jones, and K.~Andrews, ``Capacity-approaching
  protograph codes,'' \emph{IEEE Journal on Selected Areas in Communications},
  vol.~27, no.~6, pp. 876--888, 2009.

\bibitem{Divsalar}
D.~Divsalar, S.~Dolinar, and C.~Jones, ``Short protograph-based {LDPC} codes,''
  in \emph{IEEE MILCOM 2007}, 2007, pp. 1--6.

\bibitem{MacKayDist}
\BIBentryALTinterwordspacing
D.~J.~C. MacKay. (2008) {Source code for approximating the MinDist problem of
  LDPC codes}. [Online]. Available:
  \url{http://www.inference.eng.cam.ac.uk/mackay/MINDIST\_ECC.html}
\BIBentrySTDinterwordspacing

\end{thebibliography}

\end{document}